\documentclass[aps,prl,twocolumn,superscriptaddress,showpacs,longbibliography]{revtex4-2}

\usepackage{amsmath, amssymb,wasysym,graphicx}
\usepackage{appendix}
\usepackage[ruled]{algorithm2e}

\usepackage[utf8]{inputenc}
\usepackage[T1]{fontenc}
\usepackage{xcolor}

\usepackage{diagbox}
\usepackage{tabularx}

\IfFileExists{newtxtext.sty}
{\usepackage{newtxtext,newtxmath}}
{\IfFileExists{stix.sty}
	{\usepackage{stix}}
	{\IfFileExists{mathptmx.sty}
		{\usepackage{mathptmx}}{} } }

\usepackage{textcomp}

\usepackage{bm}

\usepackage{booktabs,siunitx}

\usepackage{color}
\definecolor{LinkColor}{rgb}{0.256,0.439,0.588}
\usepackage{hyperref}
\hypersetup{
	pdfauthor={good guys},
	pdftitle={good title},
	colorlinks=true,
	citecolor=LinkColor,
	linkcolor=LinkColor,
	urlcolor=LinkColor
}

\newcommand{\bra}[1]{\langle#1\rvert}
\newcommand{\ket}[1]{\lvert#1\rangle}

\def\bk{{\mathbf{k}}}

\def\bK{{\mathbf{K}}}

\def\bq{{\mathbf{q}}}
\def\bQ{{\mathbf{Q}}}

\def\bG{{\mathbf{G}}}

\usepackage{multirow}

\begin{document}
\title{Evolution from quantum anomalous Hall insulator to heavy-fermion semimetal \\in magic-angle twisted bilayer graphene}

\author{Cheng Huang}\affiliation{Department of Physics and HKU-UCAS Joint Institute of Theoretical and Computational Physics, The University of Hong Kong, Pokfulam Road, Hong Kong SAR, China}

\author{Xu Zhang}\affiliation{Department of Physics and HKU-UCAS Joint Institute of Theoretical and Computational Physics, The University of Hong Kong, Pokfulam Road, Hong Kong SAR, China}

\author{Gaopei Pan}\affiliation{Beijing National Laboratory for Condensed Matter Physics and Institute of Physics, Chinese Academy of Sciences, Beijing 100190, China}\affiliation{School of Physical Sciences, University of Chinese Academy of Sciences, Beijing 100049, China}

\author{Heqiu Li}\affiliation{Department of Physics, University of Toronto, Toronto, Ontario M5S 1A7, Canada}

\author{Kai Sun}\email{sunkai@umich.edu}\affiliation{Department of Physics, University of Michigan, Ann Arbor, Michigan 48109, USA}

\author{Xi Dai}\email{daix@ust.hk}\affiliation{Department of Physics, Hong Kong University of Science and Technology, Kowloon, Hong Kong SAR, China}

\author{Zi Yang Meng}\email{zymeng@hku.hk}\affiliation{Department of Physics and HKU-UCAS Joint Institute of Theoretical and Computational Physics, The University of Hong Kong, Pokfulam Road, Hong Kong SAR, China}

\begin{abstract}
The ground states of twisted bilayer graphene (TBG) at chiral and flat-band limit with integer fillings are known from exact solutions
, while their dynamical and thermodynamical properties are revealed by unbiased quantum Monte Carlo (QMC) simulations
. However, to elucidate experimental observations of correlated metallic, insulating and superconducting states and their transitions, investigations on realistic, or non-chiral cases are vital. Here we employ momentum-space QMC method to investigate the evolution of correlated states in magic-angle TBG away from chiral limit at charge neutrality with polarized spin/valley, which approximates to an experimental case with filling factor $\nu=-3$
. We find that the ground state evolves from quantum anomalous Hall insulator into an intriguing correlated semi-metallic state possessing heavy-fermion features as AA hopping strength reaches experimental values. Such a state resembles the recently proposed heavy-fermion representations with localized electrons residing at AA stacking regions and delocalized electrons itinerating via AB/BA stacking regions
. The spectral signatures of the localized and itinerant electrons in the heavy-fermion semimetal phase are revealed, with the connection to experimental results being discussed.
\end{abstract}
\date{\today}
\maketitle

\noindent{\textcolor{blue}{\it Introduction.}---}  
Flat bands of non-interacting magic-angle twisted bilayer graphene (TBG)  can be readily characterized by tight-binding~\cite{tramblyLocalization2010,tramblyNumerical2012} and continuous Bistritzer-MacDonald~\cite{rafiMoire2011} models. Yet, in such deceptively simple systems, it is the interplay of long-range Coulomb interaction, quantum metric of flat-band wavefunction and the extremely high tunability, by twisting angles, gating and dielectric environment, that gives rise to a plethora of novel quantum states including correlated insulator ~\cite{caoCorrelated2018,caoTunable2020,polshynLarge2019,liuTuning2021,xie2019spectroscopic,choi2019electronic,nuckolls2020strongly,
saito2021hofstadter,das2021symmetry,wu2021chern,potaszExact2021}, unconventional metal with linear-$T$ resistivity,  superconductor and beyond~\cite{jaoui2022quantum,caoUnconventional2018,cao2021nematicity,yankowitz2019tuning,diez2021magnetic,di2022revealing},  the underlying principles and mechanisms of which are still under intensive investigations.

Among these efforts, exact solutions~\cite{lianTwisted2021tbg4,xieTwisted2021,bultinck2020ground,potaszExact2021,wilhelmInterplay2021}, mean-field analyses~\cite{xieNature2020,bultinck2020ground,YiZhang2020,liuTheories2021,hejaziHybrid2021,xiePhase2023,linSymmetry2020,kwanDomain2021,dattaHeavy2023}, density matrix renormalization group (DMRG) investigations~\cite{kangNon-Abelian2020,soejimaEfficient2020}, and momentum-space quantum Monte Carlo (QMC) simulations~\cite{zhangMomentum2021,panDynamical2022,hofmannFermionic2022,panThermodynamic2023,zhangPolynomial2023} reveal that the TBG model of the first magic angle ($1.08^\circ$) with Coulomb interaction projected onto the flat bands possesses a quantum anomalous Hall (QAH) ground state with Chern number $C = 1$, at chiral (AA stacking hopping $u_0 = 0$) and flat-band limit (ignoring kinetic energy) with filling factor $\nu=\pm3$ (which means one of the eight spin/valley degenerate bands are occupied for $\nu=-3$, or seven considering particle-hole symmetry for $\nu=3$). For the same system away from chiral limit, which means $u_0>0$ (can be varied experimentally by twisting angle, temperature, strain, magnetic field, or pressure, etc~\cite{caoUnconventional2018,yankowitz2019tuning}), a DMRG investigation suggests its ground state remains Chern number  polarized until $u_0/u_1\approx0.8$ (AB/BA stacking hopping $u_1\approx110$ meV~\cite{kuzmenkoDetermination2009,andrei2020graphene,xu2020correlated}), beyond which competing states such as gapless $C_2\mathcal{T}$ nematic and gapped $C_2\mathcal{T}$ stripe phases emerge~\cite{kangNon-Abelian2020}. Nevertheless, another DMRG work indicates the ground state a nematic $C_2\mathcal{T}$ semimetal when $u_0/u_1 > 0.8$~\cite{soejimaEfficient2020}, while a recent Hatree-Fock mean-field study suggests the corresponding state to be insulating and monolayer charge density modulated with $C_2\mathcal{T}$ symmetry~\cite{xiePhase2023}. 

Experimentally, magic-angle (hereafter magic-angle refers to the fisrt magic angle) TBG at $\nu=-3$ is found to be metallic (or semi-metallic considering carrier concentration) with $C_3$ symmetric electron localization, neighboring to a superconducting phase, while that of $\nu=3$ is semiconducting or insulating~\cite{saito2020independent,saito2021hofstadter,luSuperconductors2019,xie2019spectroscopic,choi2019electronic,stepanov2020untying,polshynLarge2019,stepanov2020untying}, when $u_1$ is around 110 meV~\cite{kuzmenkoDetermination2009,andrei2020graphene,xu2020correlated} and $u_0$ around 0.8 of $u_1$~\cite{xie2019spectroscopic,namLattice2017,koshinoMaximally2018} due to lattice relaxation. 
The discrepancies between the experimentally observations (especially for $\nu=-3$) and the above theoretical proposals are intriguing.

To understand and clarify the discrepancies between theories and experiments, in this work, we apply the unbiased momentum-space QMC simulation~\cite{zhangMomentum2021,panDynamical2022,panThermodynamic2023,hofmannFermionic2022,zhangPolynomial2023,xi2022quantum}, supplemented by exact diagonalization (ED)~\cite{liSpontaneous2021}, to systematically investigate the evolution of the correlated states of magic-angle TBG at $\nu=-3$ and away from chiral limit (increasing $u_0$), with the Coulomb interaction projected onto flat bands. Our results show the critical transition temperature ($T_\mathrm{c}$) of the QAH state decreases with increasing $u_0$, and eventually vanish at $u_0 \approx$ 90 meV, the approximate value of realistic $u_0$, where a semimetallic (SM) state probably preserving $C_3$ rotation symmetry emerges, as meanwhile the quasi-particle gap reduces to zero. This semimetallic state is gapless in the vicinity of $\Gamma$ point of the moir\'e Brilliouin zone (mBZ). In real space it has localized states at the AA stacking region while itinerant states at the AB/BA stacking regions of the moir\'e unit cell (mUC), and the localized states and itinerant states are well separated in energy -- an analog to the experiments~\cite{xie2019spectroscopic,choi2019electronic} and recently found heavy-fermion representations of the TBG system, with AA stacking regions proposed to hold states of localized flat bands ($f$ electrons) and AB/BA stacking regions bridging itinerant states ($c$ electrons)~\cite{songMagic-Angle2022,shiHeavey-fermion2022,calugaruTBG2023,huKondo2023,huSymmetric2023,chouKondo2023,dattaHeavy2023,chouScaling2023}. 

Since our model has no prior heavy-fermion representation in the first place but mainly the approximation of projecting the Coulomb interaction into the flat bands, our observation, the evolution from QAH insulator to heavy-fermion semimetal at $\nu=-3$, provides an unbiased and unifying interpretation of the ground state transition of magic-angle TBG from topological insulator (at chiral limit) 
to the SM state (non-chiral) with topological heavy-fermion features.

\begin{figure}[!h]
\centering
\includegraphics[width=0.9\linewidth]{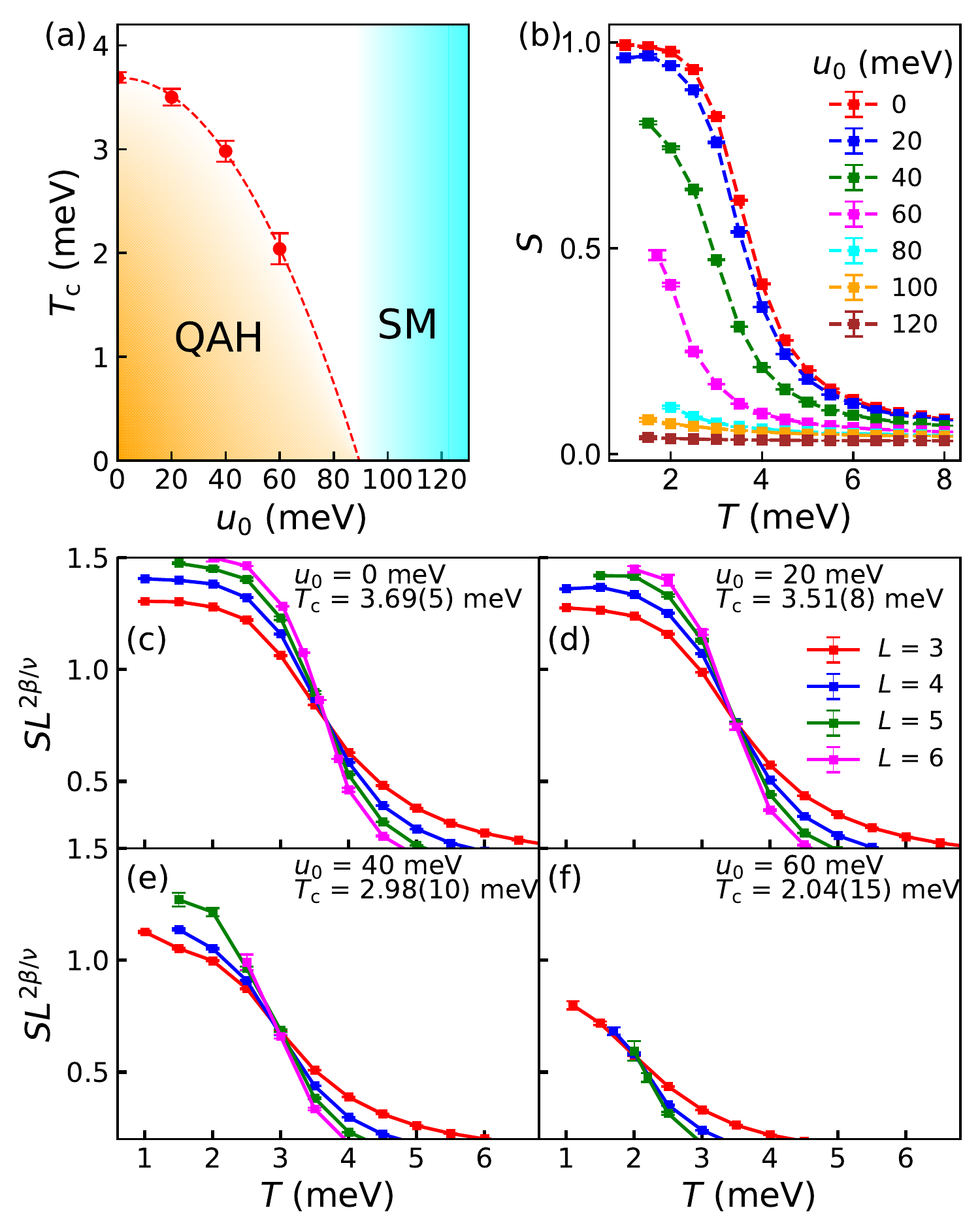}
\caption{\textbf{QAH state to heavy-fermion semimetal transition away from chiral limit in the magic-angle TBG of $\nu=-3$.} (a) The phase diagram, where $T_\mathrm{c}$ refers to the transition temperature of the QAH state, marking the time-reversal symmetry breaking and  being determinded by finite-size scaling crossings in (c)-(f), and the dased line is a fit to the data. (b) Chern number polarization, $S$, obtained from the $L=4$ system, where the simulation $T$ for each $u_0$ reaches as low temperature as possible. (c)-(f) Finite-size effect crossings with 2D Ising exponents to derive $T_\mathrm{c}$.}
\label{fig:fig1}
\end{figure}

\begin{figure}[htp!]
\centering
\includegraphics[width=\linewidth]{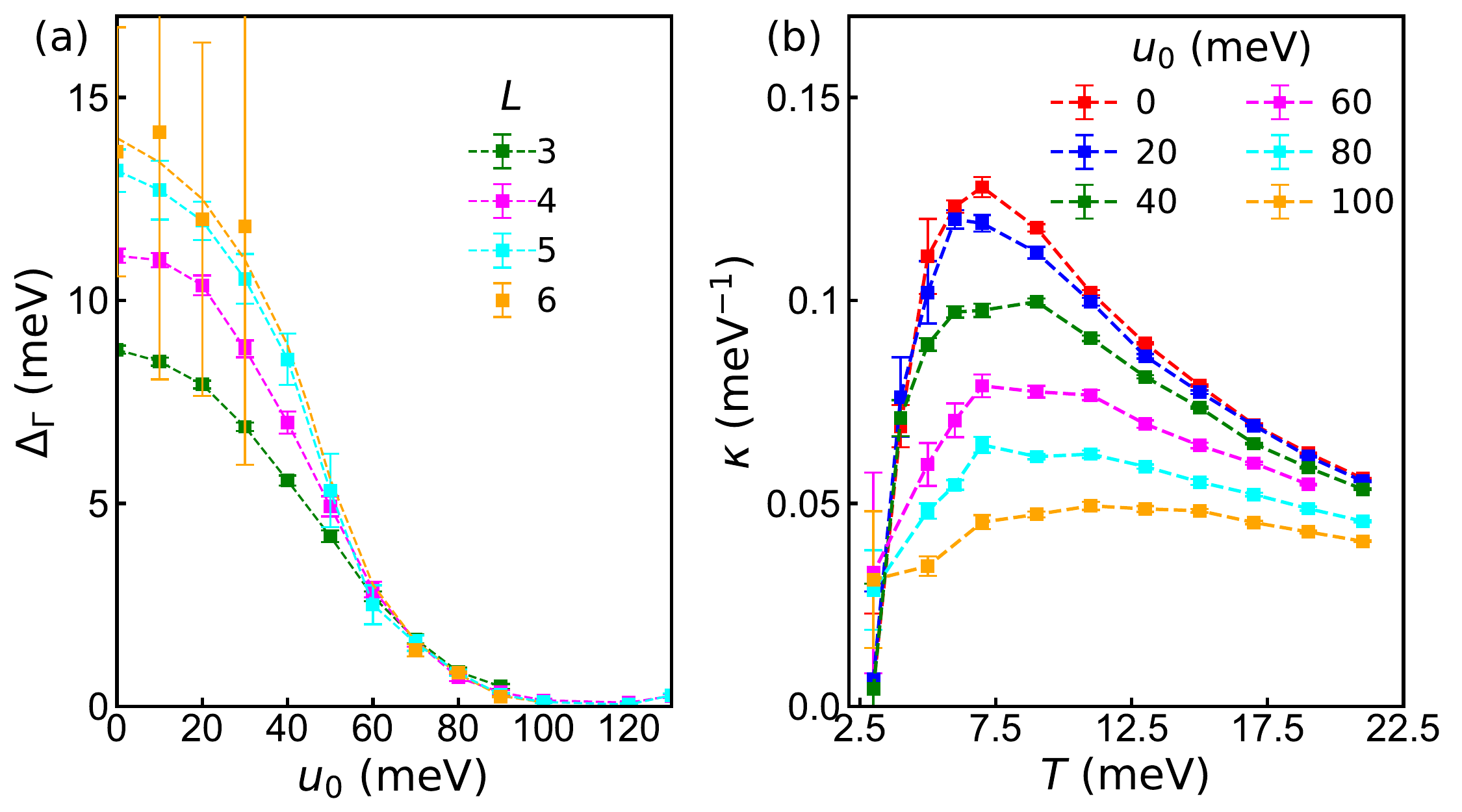}
\caption{\textbf{Quasiparticle gap of $\Gamma$, $\Delta_\Gamma$, and charge compressibility, $\kappa$, as functions of $u_0$ or $T$}. (a) The simulation temperature is 3 meV considering the sign problem and the consistency to those of ED. (b) The system size is $L=3$.
}
\label{fig:fig2}
\end{figure}

\noindent{\textcolor{blue}{\it Model and Method.}---} 
The real space moir\'e superlattice generated by TBG is shown in Fig.~\ref{fig:Lattice}. In principle, the Hamiltonian of TBG contains kenetics and interaction, $H=H_0+H_I$. The kenetic part is
\begin{equation}
H_0=\sum_{\mathbf{k},\bG,\bG',\eta,s}\mathbf{c}^\dagger_{\mathbf{k},\bG,\eta,s}H^{\eta,s}(\mathbf{k})_{\bG,\bG'}
\mathbf{c}_{\mathbf{k},\bG',\eta,s},
\label{eq:eq0}
\end{equation}
where
\begin{equation}
	\begin{aligned}
		H^{\eta,s}(\mathbf{k})_{\bG,\bG'}=&\delta_{\bk_1,\bk_2}\hbar\nu_\mathrm{F}
\left(\begin{array}{cc} 
			-(\bk_1-\bK_1^\eta) \cdot \pmb{\sigma}^\eta  &  0  \\
			0  & - (\bk_1-\bK_2^\eta) \cdot \pmb{\sigma}^\eta
		\end{array}\right) \\
		&+\left(\begin{array}{cc} 
			0  &  T^\eta_1  \\ 
			{T^\eta_2}^\dagger & 0
		\end{array}\right)
	\end{aligned}
	\label{eq:eq1}
\end{equation}
is the BM model~\cite{rafiMoire2011} with
\begin{equation}
\begin{aligned}
T^\eta_l=&\left(\begin{array}{cc}
u_0 & u_1\\
u_1 & u_0\end{array}\right)\delta_{\bk_1,\bk_2}
+\left(\begin{array}{cc}
u_0 & u_1\mathrm{e}^{-i\frac{2\pi}{3}\eta}\\
u_1\mathrm{e}^{i\frac{2\pi}{3}\eta} & u_0\end{array}\right)\delta_{\bk_1,\bk_2+(-1)^l\eta\bG_1}\\
&+\left(\begin{array}{cc}
u_0 & u_1\mathrm{e}^{i\frac{2\pi}{3}\eta}\\
u_1\mathrm{e}^{-i\frac{2\pi}{3}\eta} & u_0\end{array}\right)\delta_{\bk_1,\bk_2+(-1)^l\eta\left(\bG_1+\bG_2\right)}.
\end{aligned}
\end{equation}
Here $\bk$ is a momentum in mBZ, $\bG$ is a momentum difference for $\bk$, as $\bk_1=\bk+\bG$ and $\bk_2=\bk+\bG'$, to account for the contribution from extended Brillouin zones. $\bG,\bG'\in\{N_1\bG_1+N_2\bG_2\}$, the combinations of lattices of mBZ, with $N_1$ and $N_2$ being integers, and a cutoff, $|\bG|$, $|\bG'|\leq4|\bG_1|$, is applied. $\mathbf{c}^\dagger_{\mathbf{k},\bG,\eta,s}$ is a $1\times4$ vector of creation operators, $c^\dagger_{\mathbf{k},\bG,\eta,s,X}$, where $X$ is the index refering to layers and sublattices, as $X\in\{\mathrm{A}_1, \mathrm{B}_1, \mathrm{A}_2, \mathrm{B}_2\}$ for layers (1, 2) and sublattices (A, B). In addition, $s$ specifies spins ($\uparrow, \downarrow$) and $\eta=\pm$ labels the two valleys. $\nu_\mathrm{F}$ is the Fermi velocity and we set $\hbar v_\mathrm{F} /(\sqrt{3}a)=2.37745$ eV corresponding to flat bands, with $a$ the distance of nearest carbon atoms. $\pmb{\sigma}^\eta = \left(\eta\sigma_x,\sigma_y\right)$, where $\sigma_x$ and $\sigma_y$ are Pauli matrices refer to sublattices. The first part of $H^{\eta,s}(\bk)_{\bG,\bG'}$ stands for intra-layer hopping of the top (1) layer and the bottom (2) layer, respectively, and the second part stands for inter-layer hopping, which is actually the mori\'e potential. 

The interaction part is of density-density, as
\begin{equation}
H_\mathrm{I}=\frac{1}{2\Omega}\sum_{\bq\in\mathrm{mBZ},\bG}V(\bq+\bG)\delta\rho_{\bq+\bG}\delta\rho_{-\bq-\bG},
\end{equation}
where
\begin{equation}
\delta\rho_{\bq+\bG}=\sum_{\bk,\bG',\eta,s,X}
\left(c^\dagger_{\bk,\bG',\eta,s,X}c_{\bk+\bq+\bG,\bG',\eta,s,X}-\frac{\nu+4}{8}\delta_{\bq,0}\delta_{\bG,0}\right).
\end{equation}
Here $\delta\rho_\bq$ is the electron density operator with the reference to $(\nu+4)/8$, where $\nu$ is the filling parameter with $\nu = 0$ as the charge neutrality. $V(\bq)$ is the long-range screened Coulombic interaction decaying as $V(\mathbf{q})=\frac{\mathrm{e}^{2}}{2 \varepsilon} \frac{1}{|\bq|}\left(1-\mathrm{e}^{-|\bq|d}\right)$~\cite{panThermodynamic2023}, where $d/2$ is the distance between TBG and a bottom gate~\cite{liuNematic2021} (here we set $d=40$ nm) and $\varepsilon$ (=7 $\varepsilon_{0}$) is the dielectric constant. The fermion momentum transfer, $\bq+\bG$, is cut off at the amplitude of $|\bG_1|$. $\Omega=N_k|\mathbf{a}_\mathrm{M1}||\mathbf{a}_\mathrm{M2}|\sqrt{3}/2$ is the total area of mori\'e superlattice that we consider in real space, where $N_k=L\times L$ is the number of $\bk$ points sampling in mBZ and $L$ is the linear size, and $\mathbf{a}_\mathrm{M1}$ and $\mathbf{a}_\mathrm{M2}$ are the lattice vectors of mUC as shown in Fig. \ref{fig:Lattice}, with $|\mathbf{a}_{\mathrm{M1/M2}}|=\sqrt{3}a/[2\mathrm{sin}(\theta/2)]$ and $\theta=1.08^{\circ}$, the twisting angle.
 
As shown in Fig.~\ref{fig:bands} in the appendix, even in non-chiral cases ($u_0 > 0$), the two low-energy bands of $H_0$ are quite flat and isolated from the remote bands, except when $u_0 \approx u_1$ and $u_0 > 160$ meV. In these isolated flat-band cases, therefore, the low-energy physics of $H=H_0+H_\mathrm{I}$ can be well captured by the two flat bands, and $H$ can thus be projected onto them. Setting the eigenvalues and eigenvectors of $H^{\eta,s}(\bk)$ to be $\epsilon^{\eta,s}_m(\bk)$ and $\left|u_{\bk,m,\eta,s}\right>$, where $m = \pm$ is the index for the flat bands, and projecting $H$ onto the flat bands, as $c^\dagger_{\bk,m,\eta,s}=\sum_{\bG,X}c^\dagger_{\bk,\bG,\eta,s,X}\left|u_{\bk,m,\eta,s}\right>_{\bG,X}$, yields
\begin{equation}
\begin{aligned}
&H_0=\sum_{\bk,m,\eta,s}\epsilon^{\eta,s}_m(\bk)c^\dagger_{\bk,m,\eta,s}c_{\bk,m,\eta,s},\\
&H_\mathrm{I}=\frac{1}{2\Omega}\sum_{|\bq+\bG|\neq 0} V(\bq+\bG)\delta\rho_{\bq+\bG}\delta\rho_{-\bq-\bG},
\label{eq:eq2}
\end{aligned}
\end{equation}
where $\delta\rho_{\bq+\bG}=\sum_{\bk, m_{1}, m_{2},\eta,s}\lambda_{m_1,m_2,\eta,s}(\bk,\bk+\bq+\bG) \left(c_{\bk, m_{1},\eta,s}^{\dagger} c_{\bk+\bq+\bG, m_{2},\eta,s}  -\frac{\nu+4}{8} \delta_{\bq,0}\delta_{m_1,m_2}\right)$, and the form factor $\lambda_{m_{1}, m_{2},\eta,s}(\mathbf{k}, \mathbf{k}+\mathbf{q}+\bG) \equiv\left\langle u_{\bk, m_1,\eta,s} \mid u_{\bk+\bq+\bG, m_2,\eta,s}\right\rangle$. 
To efficiently simulate the system at filling $\nu=-3$, same as earlier studies~\cite{kangNon-Abelian2020,soejimaEfficient2020}, we take two half-filled flat bands of $\eta=+$ and $s= \uparrow$ into consideration due to its spin and valley polarized ground state~\cite{kangNon-Abelian2020,soejimaEfficient2020,xiePhase2023}, while other flat bands of the rest combinations of spin and valley, which remain empty, can be safely ignored in the simulations. Note that we strip the indices of valley and spin, $\eta$ and $s$, in the followings. The above $H$ is tackled by determinant QMC method, the detail of which is depicted in Appendix A. We would like to note that our simulation ingnores the kinetic term $H_0$, due to its flat dispersion, after the projection onto flat bands, thus corresponding to a perfect flat band limit. Moreover, our model tends to predict the nature of pristine TBG and does not take any heterostrain into account, and therefore the Incommensurate Kekule Spiral state found in experiments \cite{nuckolls2023quantum,Hong2022Detecting,Kwan2021Kekule,Wang2023Ground-state} will not be detected.

\begin{figure}[htp!]
	\centering
	\includegraphics[width=\linewidth]{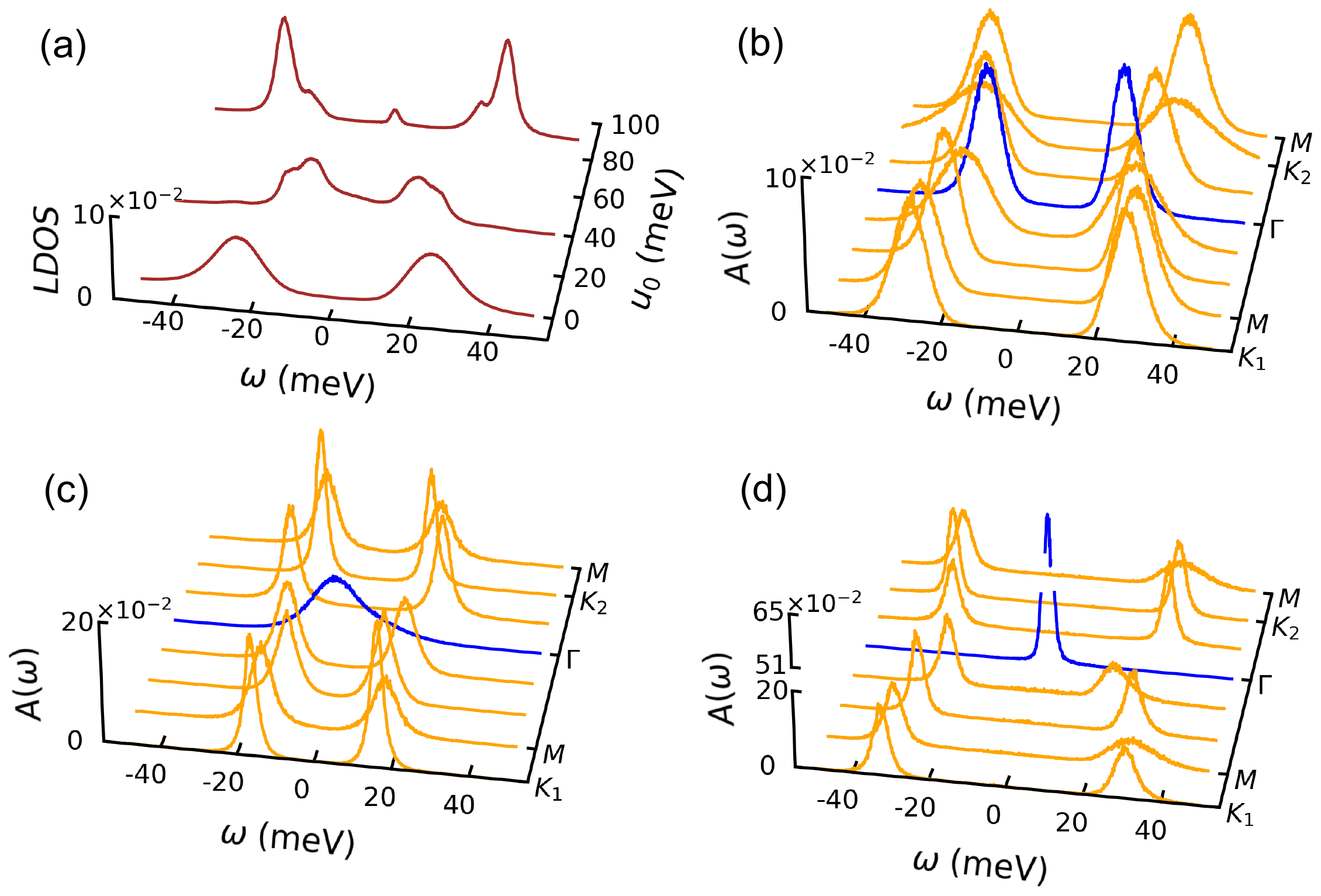}
	\caption{\textbf{$LDOS$ and quasiparticle spectra with $L=6$ and $T=3$ meV.} (a) Normalized local density of states ($LDOS$) versus energy $\omega$ and $u_0$. (b), (c), and (d) Single-particle spectra $A(\mathbf{k},\omega)$ along a high-symmetry path when $u_0=0$, $u_0=40$ meV, and $u_0=90$ meV, respectively, where the blue lines denote the spectra of $\Gamma$, while the orange lines are $A(\mathbf{k},\omega)$ of other momentum points.  Please note that the vertical axis in (d) is broken, due to the sharp peak of $\Gamma$, to make the drawing compact. Mind that $T=3$ meV is blow $T_\mathrm{c}$ when $u_0=0$ but above it when $u_0=40$ meV.}
	\label{fig:fig3}
\end{figure}

\noindent{\textcolor{blue}{\it Results.}---} 
The ground state of this system at chiral limit is found to be a Chern-number polarized QAH state~\cite{panThermodynamic2023}. To investigate if the QAH state sustains in non-chairal cases, the correlation of its QAH order is calculated. The QAH order parameter is defined as $P_\mathrm{C}/N_\bk$, where $P_\mathrm{C}=N_+-N_-$ and $N_m=\sum_\bk c^\dagger_{\bk,m}c_{\bk,m}$, the occupation number operator of two flat bands. The correlation of this QAH order, or Chern-number polarization, is $S \equiv \left\langle \left(N_{+}-N_{-}\right)^2\right\rangle/N_\bk^2$. 
Its $T$ and $u_0$ dependence for $L=4$ system is shown in Fig.~\ref{fig:fig1} (b). For specific $u_0$, $S$ increases with decreasing $T$, which manifests spontaneous time-reversal symmetry breaking and a QAH ground state (a topological insulator) at low temperature~\cite{panThermodynamic2023,chenRealization2021,linExciton2022}. For each fixed $u_0$, finite-size scaling is carried out to derive the $T_\mathrm{c}$ of the spontaneous time-reversal symmetry breaking of the QAH state, as shown in Fig.~\ref{fig:fig1} (c)-(f), where the 2D Ising exponents $\beta=1/8$ and $\nu=1$ for the scaling form $S(T,L)L^{2\beta/\nu}= f((T-T_\mathrm{c}) L^{1/\nu})$, with $f$ being the scaling function, are used. The obtained $T_\mathrm{c}$ reduces with increasing $u_0$, as shown in Fig.~\ref{fig:fig1} (a), and it can be extrapolated that, when $u_0 \approx 90$ meV, $T_\mathrm{c}$ is zero within our resolution as indicated by the dashed fitting line with an error of 5 meV.

Such vanishing of the QAH phase as a function of $u_0$ is consistent with the evolution of the quasiparticle gap of $\Gamma$, $\Delta_\Gamma$, with different system sizes ($N_k$) at the temperature 3 meV, as shown in Fig.~\ref{fig:fig2} (a). The counterparts from ED also yield the same tendency, as shown in Fig. \ref{fig:GapwithED}. The values of $\Delta_\Gamma$ are obtained from fitting imaginary time ($\tau$) Green's functions (GFs) to a range with linear $\tau$-dependence of logarithmic GFs, see Figs.~\ref{fig:G_Gamma} and \ref{fig:Gaps} in the appendix. As a system should be gapless if there is at least one kind of gapless quasiparticles, therefore, the marginal and converged values of $\Delta_\Gamma$ of different system sizes indicate the system gapless when $u_0>0.8u_1$, as shown in Fig. \ref{fig:fig2} (a) and Fig. \ref{fig:GapwithED}, although quasiparticle gaps of other $\bk$ points of mBZ remain significant even for large $u_0$, as shown in Fig.~\ref{fig:Gaps}, suggesting an SM phase. In addition, as also depicted in Fig.~\ref{fig:fig2} (a), due to finite-size effect, the value of $\Delta_\Gamma$ with a small $u_0$ increases but converges as the system size enlarges. It is found as well that the values of $T_\mathrm{c}$ are much smaller than those of the quasiparticle gaps, due to particle-hole excitons with smaller energies above the QAH phase, which restore the broken time-reversal symmetry above $T_\mathrm{c}$ and has been evidenced in our previous works~\cite{linExciton2022,panThermodynamic2023}. Moreover, the spectra (will be discussed below) of quasiparticles of the Dirac points in mBZ, $K_1$ and $K_2$ (see Fig. \ref{fig:Lattice} for their positions), from the 6$\times$6 system are always and well gapped through the $u_0$ range investigated, as shown in~Fig. \ref{fig:fig3} (b)-(d), which thus prove that the Dirac points do not play key roles in the transition towards the SM phase.

The charge compressibility, $\kappa=d\langle N\rangle/d\mu/n_0/N_k=\beta\left(\langle N^2\rangle-\langle N\rangle^2\right)/n_0/N_k$, where $N=\sum_{m}N_m$ and $n_0$ is the particle density, is also calculated to to evidence the transition from the QAH state to the SM state, whose values are shown in Fig. \ref{fig:fig2} (b). The vanishing of $\kappa$ indicates the state incompressible and insulating, and compressibly conducting otherwise. Therefore, a small $u_0$ yields an insulating ground state, which is the QAH phase, while a large $u_0$ leads to a conducting ground state, corresponding to the SM phase. Such that the Chern-number polorization, the quasiparticle gap, and the charge compressibility yield the same physics.

To further reveal the evolution from the QAH phase to the SM phase with increasing $u_0$, we derive the quasiparticle spectra, $A(\mathbf{k},\omega)$, and the correspoding local density of state, $LDOS=\sum_\bk A(\bk,\omega)/N_\bk$, from the stochastic analytic continuation scheme~\cite{Sandvik1998Stochastic,shao2023progress}, the results of $L=6$ system at $T=3$ meV of which are shown in Fig.~\ref{fig:fig3}. As is consistent to $T_\mathrm{c}$ of the QAH phase at a specific $u_0$, $A(\mathbf{k},\omega)$ is gapped for all $\bk$-points when $T$ is below $T_\mathrm{c}$ as shown in Fig.~\ref{fig:fig3} (b), and gapless around $\Gamma$ when $T$ is slightly above $T_\mathrm{c}$ as shown in Fig.~\ref{fig:fig3} (c), suggesting the sate above QAH a semimetal and the quasiparticles high energy exitons. Therefore, $T_\mathrm{c}$s set the phases boundary between QAH state and SM, indicating again the ground state of magic-angle TBG is semimetallic when $u_0>0.8u_1$, as being further evidenced by Fig.~\ref{fig:fig3} (d), which is consistent to the experimentally found SM phases~\cite{saito2020independent,luSuperconductors2019,stepanov2020untying}. The gap of $A(\Gamma,\omega)$ is the least and gradually decreases with growing $u_0$ (Fig.~\ref{fig:fig3} (b), (c) and (d), where two peaks gradually merge into a single one. The gap totally closes when $u_0=90$ meV (Fig.~\ref{fig:fig3} (d)). The $LDOS$ remains gapped except a peak around Fermi level arises as $u_0$ increases (Fig.~\ref{fig:fig3} (a)). This peak comes from the closing of $\Delta_\Gamma$, as shown in Fig.~\ref{fig:fig3} (c) and (d), while the spectra of other quasiparticles remain well gapped. 

\begin{figure}[!h]
	\centering
	\includegraphics[width=\linewidth]{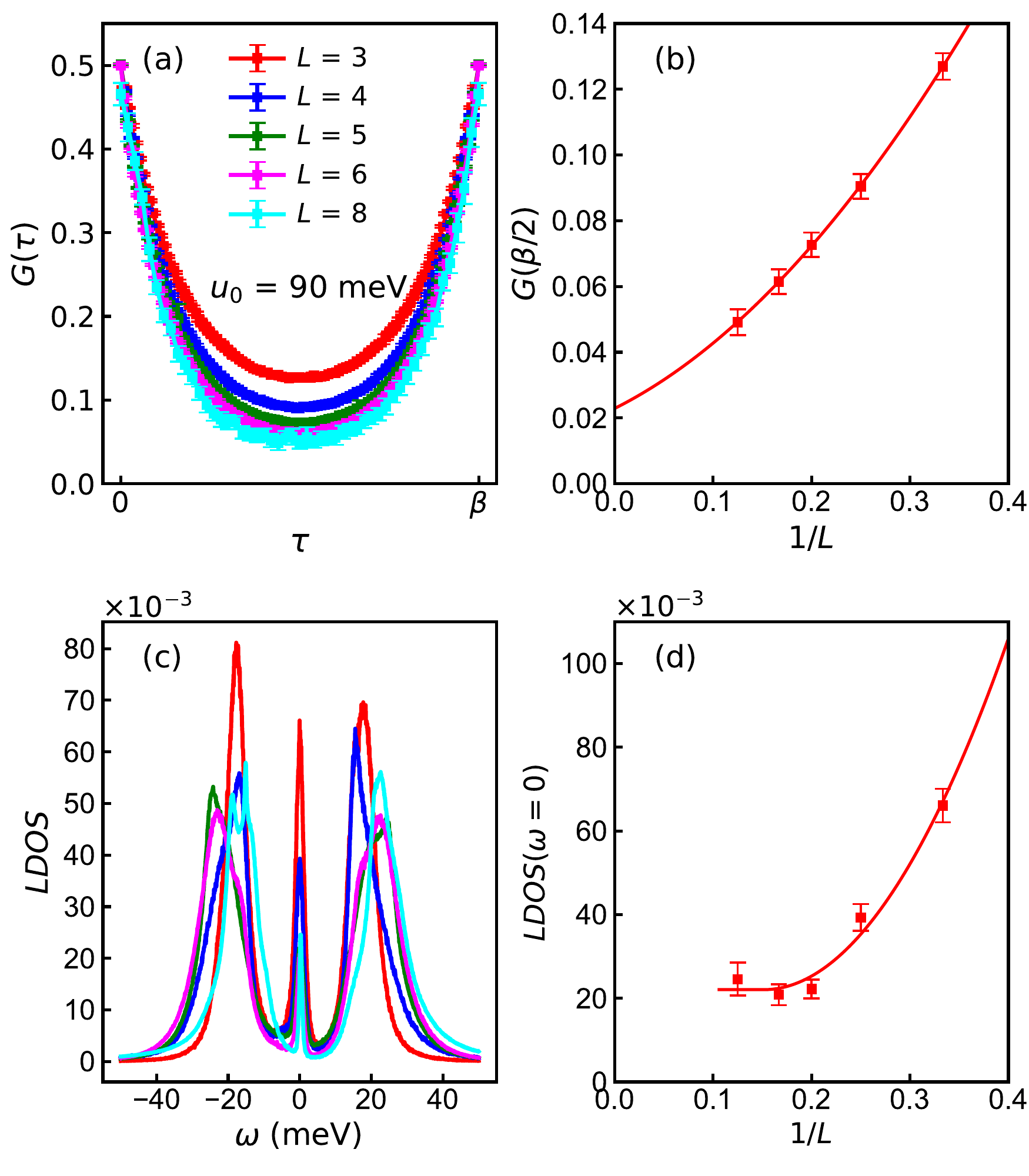}
	\caption{The decaying of average Green's function (a) and the reducing of the peak at Fermi level of $LDOS$ (c) with the system size up to $L=8$ when $u_0=90$ meV. The extrapolations of $G(\beta/2)$ and the intensity of the peak of $LDOS$ at Fermi level to larger sizes are shown in (b) and (d), respectively, where the fitting is quadratic due to that it is versus $1/L$, except the horizontal part in (d).}
	\label{fig:fig4}
\end{figure}

Average Green's function summing over $\bk$ points and $LDOS$ with different linear sizes are demonstrated in Fig.~\ref{fig:fig4}, where $u_0=90$ meV. The convergence of decaying of $G(\tau)$ can be seen, and $G(\beta/2)$ reduces probably to a minimum but finite value at the large $L$ limit, as shown in (a) and (b) respectively, suggesting that there are finite electrons at the Fermi level. Meanwhile, the peak of $LDOS$ at Fermi level decreases probably to a minimum but finite value as the system size is larger than 6, as depicted in Fig.~\ref{fig:fig4} (d). The probable non-zero extropaltions suggest quadratic band touching of the semimetallic phase of the magic-angle TBG when $u_0=90$ meV. However, a linear fit extrapolating to 0, indicating degenerate Dirac cone touching, cannot be totally ruled out in both Fig.~\ref{fig:fig4} (b) and (d). To further verify that the probable semimetallic state is of quadratic band touching, here we check if the $C_3$ rotation symmetry is broken by scrutinizing the correlation of a nematic order parameter, which is defined as $\mathcal{N}=\frac{1}{N_k}\sum_{k}c^\dagger_{\bk,+}c_{\bk,-}$. The correlation of $\mathcal{N}$, $S_\mathcal{N}=\langle \mathcal{N}^\dagger \mathcal{N}\rangle$, is calculated and illustrated in Fig.~\ref{fig:S_C}, as the system is in the SM phase. Since the value of $S_\mathcal{N}$ extrapolates to zero towards the thermodynamic limit, the $C_3$ rotation symmetry is probably not broken. Therefore, our found SM phase is probably not nematic, enhancing the probability of the SM state being a quadratic band touching semimetal.

As degenerate Dirac cone touching cannot be ruled out, we argue that it is less probable in our case. Based on the Nielsen–Ninomiya theorem of fermion doubling~\cite{nielsen1981no}, Dirac cones shall appear in pairs. The only way obeying the fermion doubling theorem is to have two degenerate Dirac cones
at $\Gamma$. However, considering that we only have one valley/spin degree of freedom
(no symmetry to enforce two-fold degeneracy), there is no obvious reason why the system would stabilize
such a two-fold degenerate energy spectrum. Nevertheless, it should be emphasized that this is not impossible, and it would imply that this semimetal phase is highly unconventional. Aside from the unconventional pathway towards Dirac semimetal, there is another less exotic possibility producing Dirac points: nematic Dirac semimetal. The system has a pair of Dirac points near $\Gamma$, one at a momentum point ${\bf q}$ and the other at ${\bf -q}$. However, such a scenario requires the breaking of rotational symmetry, which is probably preserved in our case, as indicated by $S_\mathcal{N}$. For quadratic band touching, as there is no constraint of doubling theorem, it is allowed to have a single quadratic band crossing at $\Gamma$.

The closing of the quasiparitcle gap of $\Gamma$ is consistent with the recently proposed heavy-fermion picture of magic-angle TBG~\cite{songMagic-Angle2022,shiHeavey-fermion2022,calugaruTBG2023,huKondo2023,huSymmetric2023,chouKondo2023,dattaHeavy2023,chouScaling2023}. In the heavy-fermion representations, the flat bands are mostly composed of localized electrons, except for the tiny vicinity of $\Gamma$, which is mainly contributed by delocalized electronic states. In the flat-band limit, strong interactions between localized electrons ought to induce energy gaps in the quasiparticle spectra. However, these interaction-induced gaps are expected to be much smaller in the narrow surrounding of $\Gamma$, since the fermion states there are delocalized with weaker interaction, and it is indeed so at chiral limit and when $u_0=40$ meV, as shown in Fig.~\ref{fig:fig3} (b) and (c). Moreover, the interactions among fermions around $\Gamma$ can be further weaker as the system deviates from chiral limit, and the corresponding  quasiparitcle gaps decline and eventually close when $u_0$ goes beyond a threshold value ($u_0\approx0.8u_1$ judged from $T_\mathrm{c}$), as also illustrated by Fig.~\ref{fig:fig3} (d).

\begin{figure}[htp!]
	\centering
	\includegraphics[width=\linewidth]{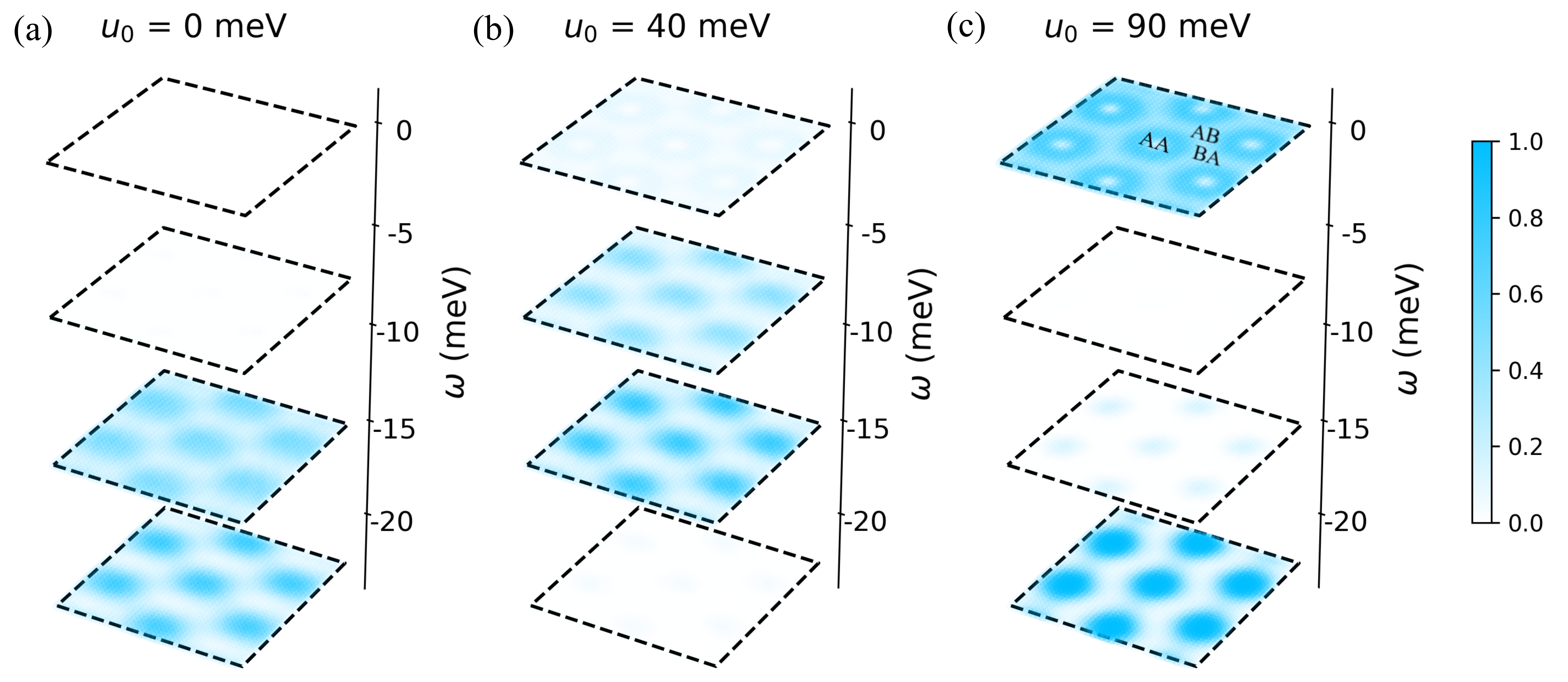}
	\caption{\textbf{Real-space spectra $A(r,\omega)$ at various $u_0$ and $\omega$ when $T=3$ meV.} Here $u_0$ equals to (a) 0 meV, (b) 40 meV, and (c) 90 meV respectively, while $\omega$ is chosed to be 0, 7.5, 15, and 22.5 meV in each figure. The intensity of blue stands for the value of $A(r,\omega)$, with a color bar locating at the right of (c). Positions of AA, AB, and BA stacking areas are labeled in (c).
	}
	\label{fig:fig5}
\end{figure}

To further scrutinize the heavy-fermion features of the SM phase, here we directly probe the localized and delocalized electron states via real-space spectra by projecting $A(\bk,\omega)$ of the $L=6$ system at $T=3$ meV with the real-space wavefunction $\psi_\bk(\mathbf{r},X)$ of $H_0$, $A(r,\omega)=\sum_{\bk,X}\psi^\dagger_\bk(\mathbf{r},X)A(\bk,\omega)\psi_\bk(\mathbf{r},X)$, as a function of $u_0$ and shown in Fig.~\ref{fig:fig5}. It can be clearly seen that the density of low energy electrons ($\omega=0$ meV, first row of panels (a) - (c)) increases in AB/BA stacking regions (or is lifted as the maximum values are around AA stacking center, where however the minimum values locate), marking the emergence of the SM state. Meanwhile, the localized electrons ($\omega=15, 22.5$ meV, bottom two rows of panels (a) - (c)) always and further concentrate at AA centers as $u_0$ grows, which is consistent with the heavy-fermion picture~\cite{songMagic-Angle2022,shiHeavey-fermion2022} and experiments~\cite{xie2019spectroscopic,choi2019electronic}. Furthermore, the density of electrons with intermediate energy ($\omega=7.5$ meV, second row in panel (c)) manifests a minimum value, separating the itinerant and localized eletrons when $u_0=90$ meV. These observations suggest that the realistic (when $u_0\approx0.8u_1$) magic-angle TBG of $\nu=-3$ can be preliminary viewed as a heavy-fermion system. Besides, the itinerant electrons of $A(r,\omega=0)$ in panel (c) come from $A(\Gamma,\omega=0)$ in Fig.~\ref{fig:fig3} (d).

\noindent{\textcolor{blue}{\it Discussions.}---} 
In this study, an evolution from the Chern insulator to a heavy-fermion semimetal probably preserving $C_3$ symmetry is found for $\nu=-3$ filling magic-angle TBG by increasing AA hopping strength, $u_0$, which can be experimentally altered. The ground state phase transition happens at $u_0\approx0.8u_1$ or 90 meV, where the Chern-number polorization and the quasiparticle gap of $\Gamma$ vanish, the system becomes compressible, and a peak in $LDOS$ at Fermi level emerges due to the gap's closing. The intensity of this peak rises as $u_0$ increases for a specific system size, but reduces probably to a minimum but finite value as $L$ enlarges for a specific $u_0$, which manifests a semimetal state as the gaps of other momenta remain significant. The momentum and real-space distributions of the localized electrons are well separated from those of itinerant electrons, which are of the $\Gamma$ point. The nature of the found semimetal state implies that the magic-angle TBG can be preliminarily, based on the numerics here, treated as a heavy-fermion system when $\nu=-3$. 
Our findings shed new lights on the correlated nature of TBG and provide critical information for understanding this fascinating system. 

{\it Acknowledgments}\,---\,We thank DongKeun Ki and Wang Yao for useful discussions. CH, XZ, GPP and ZYM acknowledge the support from the Research Grants Council (RGC) of Hong Kong SAR of China (Projects No. 17302223, No. 17301721, No. AoE/P-701/20, No. 17309822,  and No. C7037-22GF), the ANR/RGC Joint Research Scheme sponsored by RGC of Hong Kong and French National Research Agency (Project No. A\_HKU703/22), the Strategic Priority Research Program of Chinese Academy of Sciences (Grant No. XDB33000000), and K. C. Wong Education Foundation (Grant No. GJTD-2020-01). We thank the HPC2021 system of Information Technology Services and the Blackbody HPC system at the Department of Physics, of the University of Hong Kong, and Beijing PARATERA Tech Co., Ltd. (https://cloud.paratera.com) for providing computational resources that have contributed to the research results in this paper.

\renewcommand{\thefigure}{S\arabic{figure}}
\setcounter{figure}{0} 

\section{Appendix A: The momentum-space determinant QMC method}
\label{sec:AA}
Due to the flatness of the lowest energy bands, as shown in Fig.~\ref{fig:bands}, $H_0$ is ignored in our simulations. The partition function thus reads $Z=\mathrm{Tr}\left(\mathrm{e}^{-\beta H_\mathrm{I}}\right)$, where $\beta=1/k_\mathrm{B}T$ and $T$ is the temperature. $Z$ can be Trotter-decoupled after spliting inverse temperature into $N_\tau$ pieces, $\beta=\triangle\tau N_\tau$, with an error of order $\triangle\tau^2$. That is
\begin{equation}
\begin{aligned}
Z&=\mathrm{Tr}\left(\mathrm{e}^{-\beta H_\mathrm{I}}\right)\\
&=\mathrm{Tr}\left(\left(\mathrm{e}^{-\triangle\tau H_\mathrm{I}}\right)^{N_\tau}\right)\\
&=\mathrm{Tr}\left(\prod^{\beta}_{\tau=\triangle\tau}\mathrm{e}^{-\triangle\tau H_\mathrm{I}}\right)+\mathcal{O}(\triangle\tau^2).
\end{aligned}
\end{equation}
Denoting $\bQ=\bq+\bG$, $H_\mathrm{I}$ can be rewritten as
\begin{equation}
\begin{aligned}
H_\mathrm{I}&=\sum_{|\bQ|\neq 0}\frac{1}{2\Omega}V(\bQ)\delta\rho_{\bQ}\delta\rho_{-\bQ}\\
&=\sum_{\mathcal{Q}}\frac{1}{2\Omega}V(\mathcal{Q})(\delta\rho_{\mathcal{Q}}\delta\rho_{-\mathcal{Q}}+\delta\rho_{-\mathcal{Q}}\delta\rho_{\mathcal{Q}})\\
&=\sum_{\mathcal{Q}}\frac{1}{4\Omega}V(\mathcal{Q})\left(\left(\delta\rho_{-\mathcal{Q}}+\delta\rho_\mathcal{Q}\right)^2-\left(\delta\rho_{-\mathcal{Q}}-\delta\rho_\mathcal{Q}\right)^2\right),
\end{aligned}
\end{equation}
where $\mathcal{Q}$ is the half of $|\bQ|\neq0$~\cite{zhangMomentum2021}. Thus, $\mathrm{e}^{-\triangle\tau H_\mathrm{I}}$ can be further decoupled to an auxiliary field by the Hubbard-Stratonovich transformation, with an error of order $\triangle\tau^4$, as
{\small
\begin{equation}
\begin{aligned}
&\mathrm{e}^{-\triangle\tau H_\mathrm{I}}=\\
&\prod_{\mathcal{Q}}\frac{1}{16}\sum_{l_{1\mathcal{Q}},l_{2\mathcal{Q}}}\gamma(l_{1\mathcal{Q}})\gamma(l_{2\mathcal{Q}})\mathrm{e}^{i\sqrt{\alpha}\eta(l_{1\mathcal{Q}})\left(\delta\rho_{-\mathcal{Q}}+\delta\rho_\mathcal{Q}\right)}\mathrm{e}^{\sqrt{\alpha}\eta(l_{2\mathcal{Q}})\left(\delta\rho_{-\mathcal{Q}}-\delta\rho_\mathcal{Q}\right)}\\
&+\mathcal{O}(\triangle\tau^4),
\end{aligned}
\end{equation}}
where $l=\pm 1,\pm 2$, is the auxilariy field, and $\gamma (\pm 1)=1+\sqrt{6}/3$, $\gamma (\pm 2)=1-\sqrt{6}/3$, $\eta (\pm 1)=\pm\sqrt{6-2\sqrt{6}}$, and $\eta (\pm 2)=\pm\sqrt{6+2\sqrt{6}}$. $\alpha=\triangle\tau V(\mathcal{Q})/4\Omega$.

The partition function can be finally written as $Z=\sum_{\{l_{\tau1\mathcal{Q}},l_{\tau2\mathcal{Q}}\}}\mathrm{Tr}U_C$, with
\begin{equation}
\begin{aligned}
&U_C=\prod^{\beta}_{\tau=\triangle\tau}\prod_{\mathcal{Q}}\\
&\frac{1}{16}\gamma(l_{\tau1\mathcal{Q}})\gamma(l_{\tau2\mathcal{Q}})\mathrm{e}^{i\sqrt{\alpha}\eta(l_{\tau1\mathcal{Q}})\left(\delta\rho_{-\mathcal{Q}}+\delta\rho_\mathcal{Q}\right)}\mathrm{e}^{\sqrt{\alpha}\eta(l_{\tau2\mathcal{Q}})\left(\delta\rho_{-\mathcal{Q}}-\delta\rho_\mathcal{Q}\right)},
\end{aligned}
\end{equation}
a configuration of $\{l_{\tau1\mathcal{Q}},l_{\tau2\mathcal{Q}}\}$ labeled by $C$. Physical quantities, $O$, can thus be obtained by sampling the auxiliary field, as
\begin{equation}
\begin{aligned}
\langle O\rangle&=\frac{\mathrm{Tr}\left(\mathrm{e}^{-\beta H_\mathrm{I}}O\right)}{Z}\\
&=\sum_C\frac{\mathrm{Tr}\left(U_CO\right)}{Z}\\
&=\sum_C\frac{\mathrm{Tr}\left(U_C\right)}{Z}\frac{\mathrm{Tr}\left(U_CO\right)}{\mathrm{Tr}\left(U_C\right)},
\end{aligned}
\end{equation}
with $\mathrm{Tr}\left(U_C\right)/Z$ or $\mathrm{Tr}\left(U_C\right)$ the sampling weight and denoted as $P_C$.

\begin{figure}[!h]
	\centering
	\includegraphics[width=0.7\linewidth]{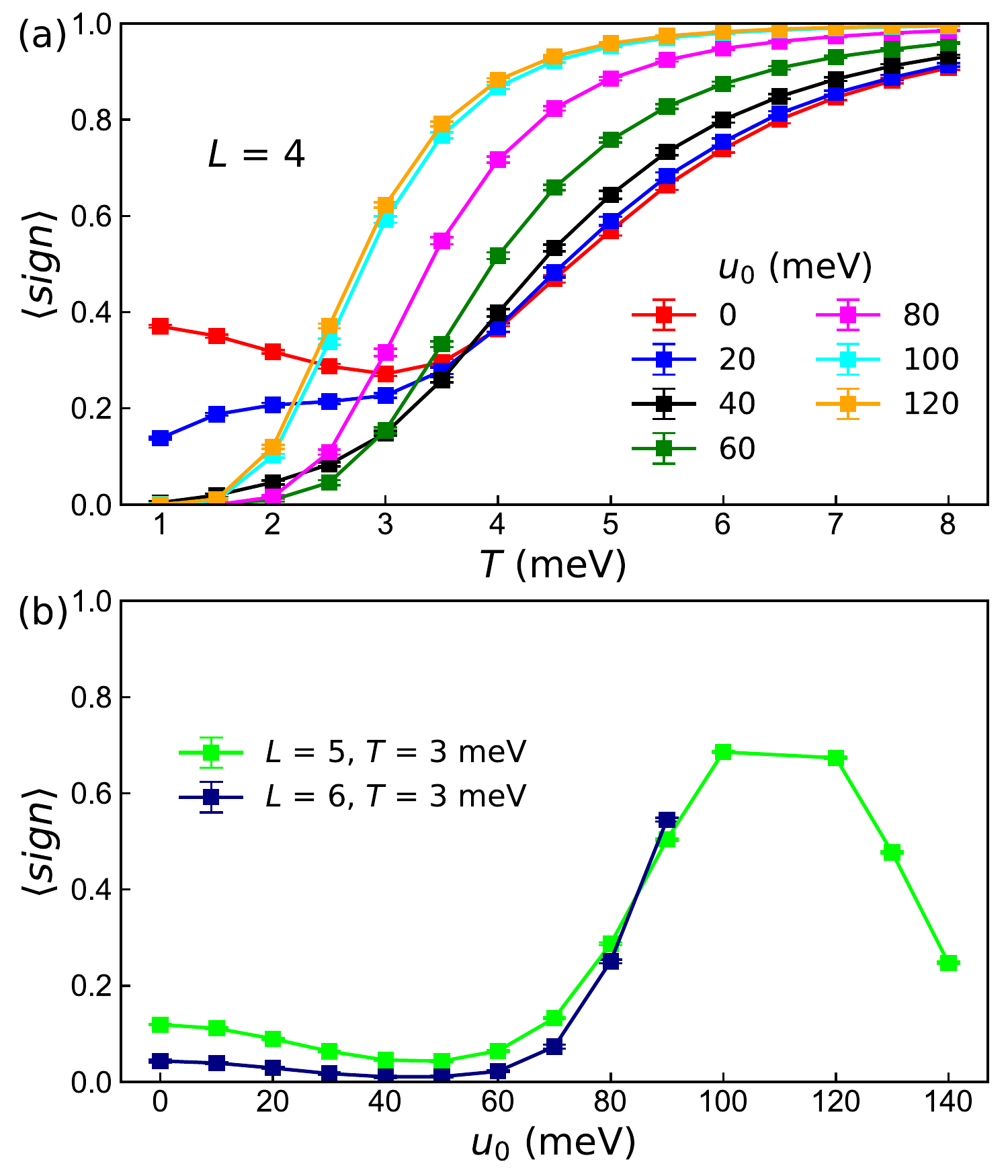}
	\caption{$\langle sign\rangle$ of magic-angle TBG with $\nu=-3$, when (a) $L=4$ varying $T$ and $u_0$, and (b) $L=5, 6$ and $T=3$ meV  varying $u_0$. $\langle sign\rangle$ reaches 0 when $u_0\geq40$ meV as $T=1$ meV in (a), while $\langle sign\rangle$ of $L=6$ reaches 0.01 when $u_0=40$ meV in (b).}
	\label{fig:Sign}
\end{figure}

However, $P_C$ of the magic-anlge TBG with $\nu=-3$ is not constantly positive, or sign-problematic, as shown in Fig.~\ref{fig:Sign}. Nevertheless, the symmetries of TBG make $P_C$ real~\cite{zhangMomentum2021} and the average of its sign, $\langle sign\rangle=\sum_Csgn(P_C)$, is always positive and ranges from 0 to 1. Here 1 means no sign problem while 0 means the simulation is valueless. It is quite intuitive how to avoid negitive weight. Denoting $O_C=\mathrm{Tr}\left(U_CO\right)/\mathrm{Tr}\left(U_C\right)$, $\langle O\rangle$ can be rewritten as
\begin{equation}
\begin{aligned}
\langle O\rangle&=\frac{\sum_C O_C P_C}{\sum_C P_C}\\
&=\frac{\sum_C O_C \mathrm{sgn}(P_C)|P_C|/ \sum_C|P_C|}{\sum_C \mathrm{sgn}(P_C)|P_C|/ \sum_C|P_C|}
\end{aligned}
\end{equation}
Thus, the numerator and the denominator can be sampling separately by $|P_C|/ \sum_C|P_C|$, with $O_C \mathrm{sgn}(P_C)$ and $\mathrm{sgn}(P_C)$ being the measured quantities, and $\langle O\rangle$ can be further obtained.

Our stable and reproducible results show that a QMC simulation can still be efficiently and reliably done when $\langle sign\rangle >0.01$ with reasonable computing resources. In the obtaining of reliable imaginary time Green's function, the simulation temperature is set to be $T=3$ meV considering the $\langle sign\rangle$ of the $L=6$ system, while the $T$ can be lowered when measuring Chern-number polarization. With the reliable imaginary time Green's function, we further employ the stochastic analytic continuation (SAC)~\cite{panDynamical2022,yanRelating2021,zhangSuperconductivity2021,zhouAmplitude2021,yanTopological2021,panThermodynamic2023} method to extract the real frequency spectra.


\section{Appendix B: Moir\'e superlattice and single-particle dispersions}
\begin{figure}[htp]
	\includegraphics[width=\linewidth]{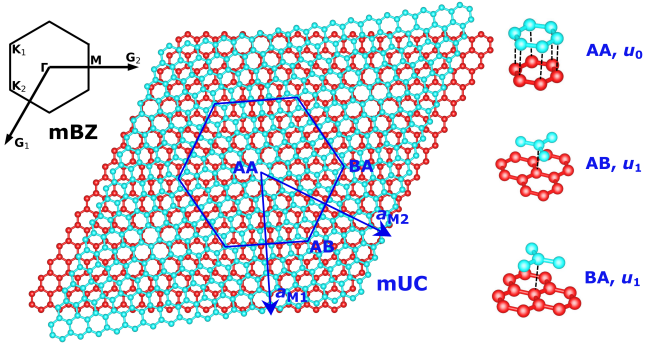}
	\caption{Moir\'e unit cell (blue hexagon) and its Brillouin zone (black hexagon), as well as three kinds of stackings and hoppings (AA, AB, and BA) of TBG. $u_0$ and $u_1$ are hopping strengths of AA- and AB/BA-stacking areas, respectively. The top (bottom) layer of TBG are denoted by cyan (red) color. $\{\mathbf{a}_\mathrm{M1},\mathbf{a}_\mathrm{M2}\}$ and $\{\mathbf{G}_1, \mathbf{G}_2\}$ are the lattice vectors of mUC and mBZ, respectively. $\Gamma$ is the origin of mBZ, and $\mathbf{K}_1$ and $\mathbf{K}_2$ are the two Dirac points of the same vellay and different layers (1, 2).  The twisted angle $\theta$ in this schematic is $10^{\circ}$, while that of the case we investigate in the main text is $1.08^{\circ}$.
}
	\label{fig:Lattice}
\end{figure}

The real space moir\'e superlattice generated by TBG is shown in Fig.~\ref{fig:Lattice}, where a mUC and its riciprocal first Brillouin zone (mBZ) are drawn. The module of the lattice vectors of mBZ is $|\bG_{1}|, |\bG_2|=8\pi\mathrm{sin}(\theta/2)/(3a)$. $\bK_1^\eta$ and $\bK_2^\eta$ are Dirac points from the top (1) and bottom (2) layers, respectively. Schematics of hopping patterns are also  shown in the right panel of Fig.~\ref{fig:Lattice}. In this work, we consider the system with the first magic angle ($\theta=1.08^\circ$).

\begin{figure}
	\centering
	\includegraphics[width=\linewidth]{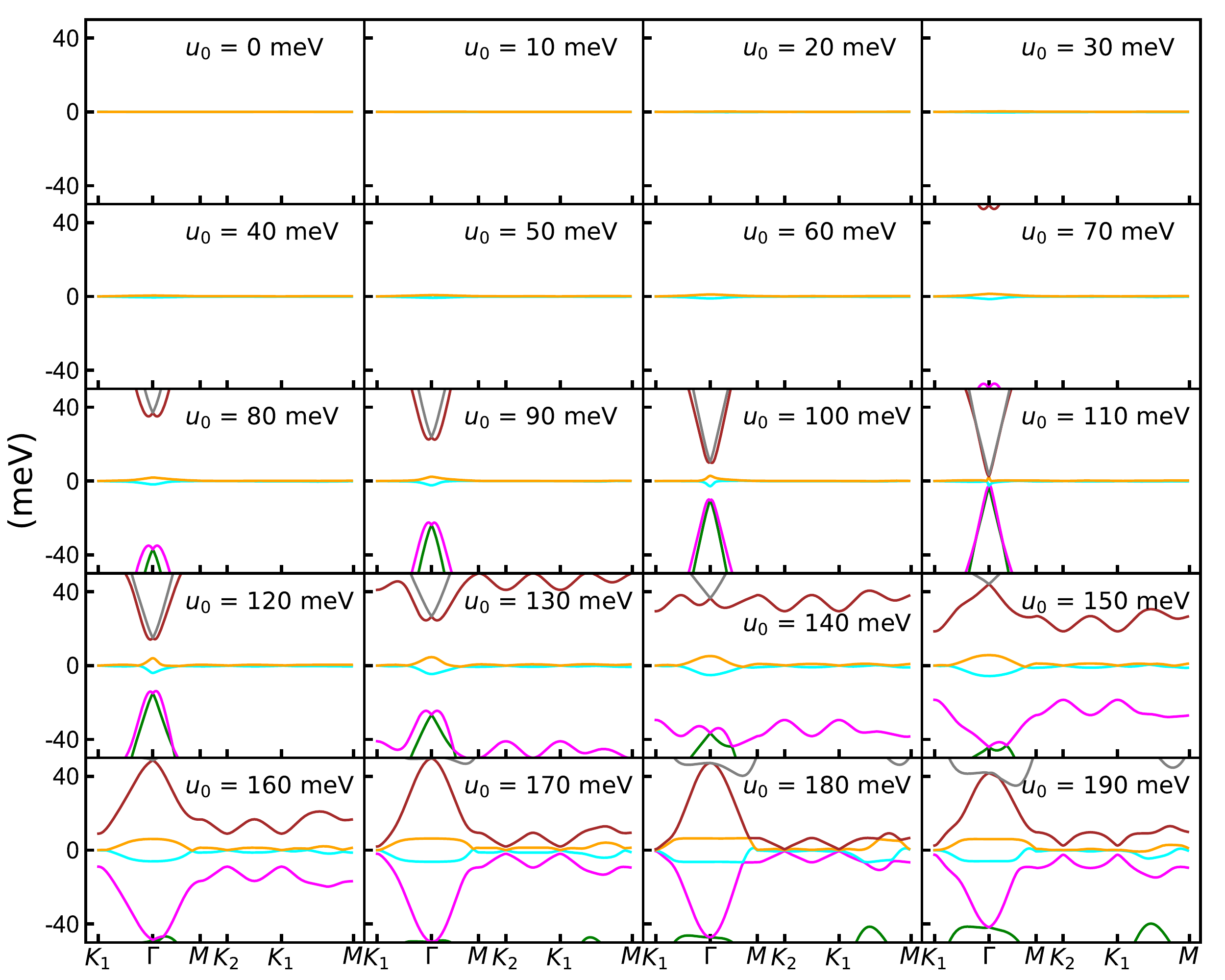}	
	\caption{Single-particle dispersions of $H_0$ at the first magic angle and along a high-symmetry line. The remote bands are beyond the scale when $u_0<70$ meV, and approches the flat bands around $\Gamma$ point as $u_0$ further increases. Note that the lowest remote bands touch the flat bands at $\Gamma$ point only when $u_0=110$ meV ($u_0=u_1$). The band width of the flat bands enlarge as $u_0$ increases, but remainds a quite small value (less than 5 meV when $u_0=90$ meV).}
	\label{fig:bands}
\end{figure}

From this moir\'e supperlattice, a momentum space single-particle Hamiltonian, $H_0$, considering intralayer hopping and interlayer hopping, can be constructed, as descriped in the main text. The dispersions of $H_0$ at the magic angle with different $u_0$ are shown in Fig.~\ref{fig:bands}, where two low energy bands are quite flat and isolated from remote bands when $u_0\leq90$ meV. It implies that the low energy physics can be well captured by these two bands. The projection of $H$ into flat bands, as depicted in the main text, makes the description of $H$ easier since only two bands are considered. However, we suppose that the projections when $u_0=110$ (meV) and $u_0>160$ (meV) fail as remote bands are too close to flat bands.

\section{Appendix C: The Calculation of Chern-number polorization}
A perturbation term, which reads
\begin{equation}
\begin{aligned}
\delta H=\left(\begin{array}{cc}
0 & 0\\
0 & \delta\sigma_\mathrm{z}\end{array}\right),
\end{aligned}
\end{equation}
with $\delta=1$ meV and $\sigma_\mathrm{z}$ being the third Pauli matrix, is added to the bottom layer of $H_0$ to break inversion and an isolated Chern bands basis $|\psi\rangle$ is obtained. This perturbation term 
breaks $C_\mathrm{2z}\mathcal{T}$ but $\mathcal{T}$, which means the spontaneous time-revesal symmetry breaking of the TBG system is still alowed. Denoting the original basis as $|\phi\rangle$, the imaginary-time Green's function in the isolated Chern bands basis can thus be related to the original basis, by $\langle\psi|G(\bk)|\psi'\rangle=\sum_{\phi,\phi'}\langle\psi|\phi\rangle\langle\phi|G(\bk)|\phi'\rangle\langle\phi'|\psi'\rangle$. Thus, the correlation function of the QAH order, or Chern-number polarization, can be calculated in the isolated Chern band basis. As this perturbation term resembles an h-BN potential, therefore the strength, $\delta$, which can be as low as 0.3 meV, is choosed to be 1 meV to minimize the relavance. This perturbation term is only added when calculating Chern-number polarization.

\section{Appendix D: Imaginary time Green's funtion and derived gaps}
\begin{figure}[!h]
	\centering
	\includegraphics[width=0.8\linewidth]{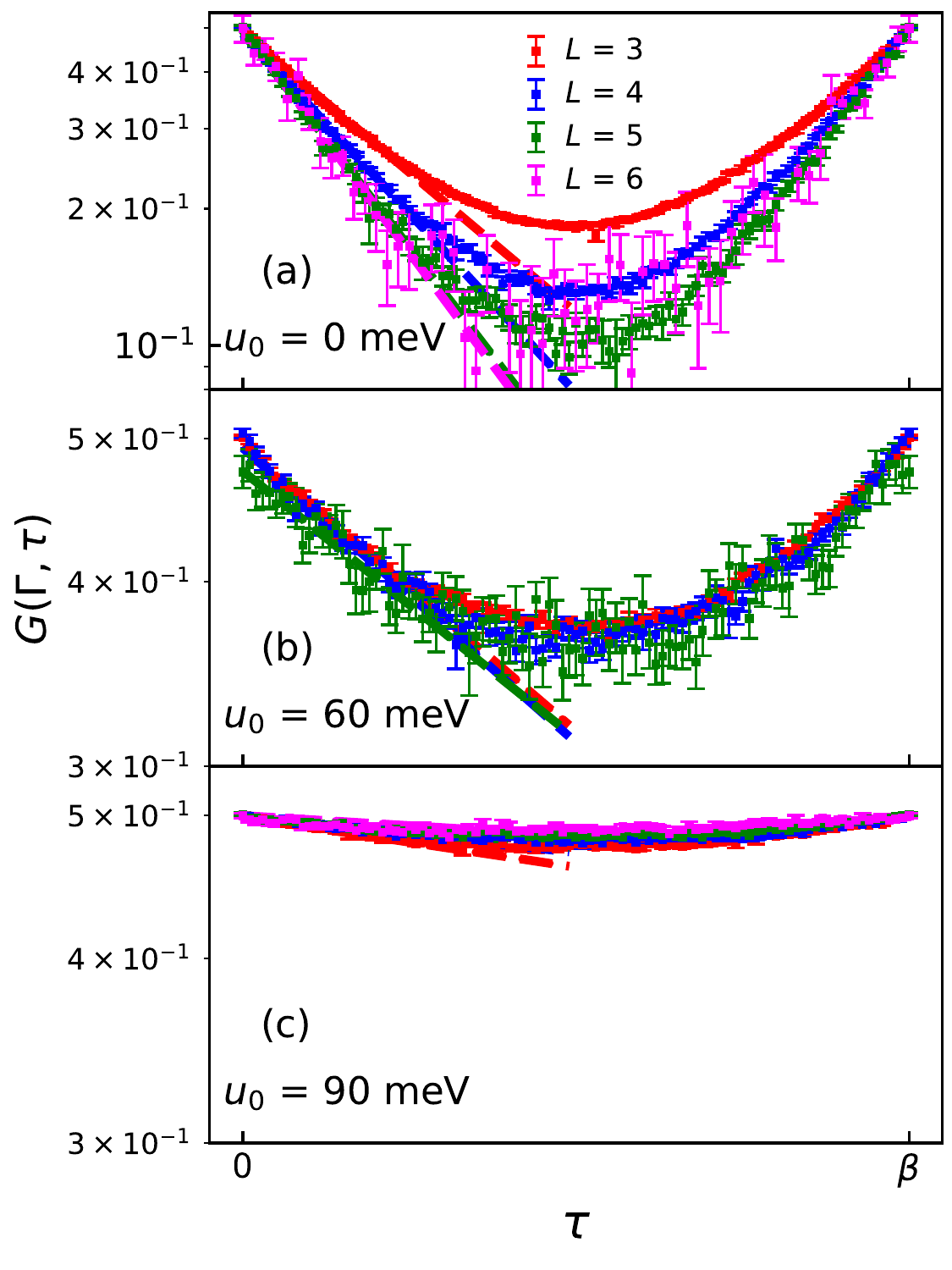}
	\caption{Decaying of imaginary time Green's function of $\Gamma$ point with $L=3, 4, 5, 6$, when (a) $u_0=0$ meV, (b) $u_0=60$ meV, and (c) $u_0=90$ meV, where the log scale is used. Mind that the vertical scales of (b) and (c) is different from that of (a). The fitting dashed lines are also shown in each panel, the slopes of which give the quasiparticle gaps in Fig.~\ref{fig:fig2} (a) and Fig.~\ref{fig:Gaps}.}
	\label{fig:G_Gamma}
\end{figure}

\begin{figure}[!h]
	\centering
	\includegraphics[width=\linewidth]{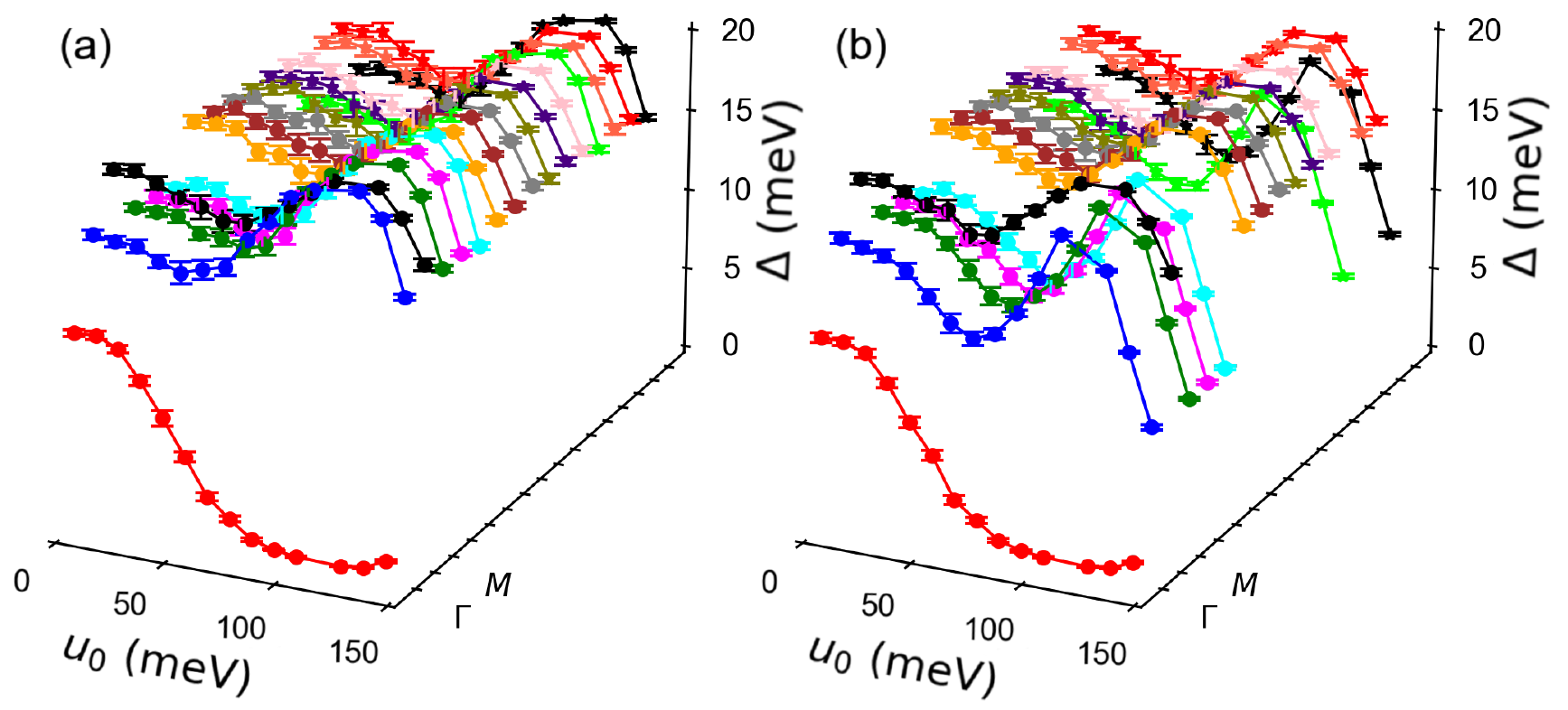}
	\caption{Quasiparticle gaps of two bands, $\pm$, (a) and (b), of $\bk$ points in mBZ with $L=4$ and varing $u_0$. Here $\Gamma$ and $M$ refer to the red circle-line and the black circle-line, respectively, and the rest circle-lines or star-lines are those of the rest momentum points. Error bars of many data points are smaller than corresponding markers' sizes. Note that the gaps of - band remain quite large values ($>8$ meV) when $u_0=150$ meV, as shown in (b).}
	\label{fig:Gaps}
\end{figure}

The decaying of $G(\Gamma,\tau)$ is depicted in Fig.~\ref{fig:G_Gamma}, where the convergence of decaying can be seen with increasing $L$. The minimum single-particle gap of each momentum can be obtained by fitting $G(\tau)=\mathrm{exp}(-\Delta\tau)/2+\mathrm{exp}\left(-\Delta(\beta-\tau)\right)/2$ to the data in a range with linear $\tau$-dependence of logarithmic GFs as shown in Fig.~\ref{fig:G_Gamma}, and the obtained gaps are demonstrated in Fig.~\ref{fig:fig2} (a) and Fig.~\ref{fig:Gaps}, where those of other $\bk$ points are also shown.

\begin{figure}[!h]
	\centering
	\includegraphics[width=0.6\linewidth]{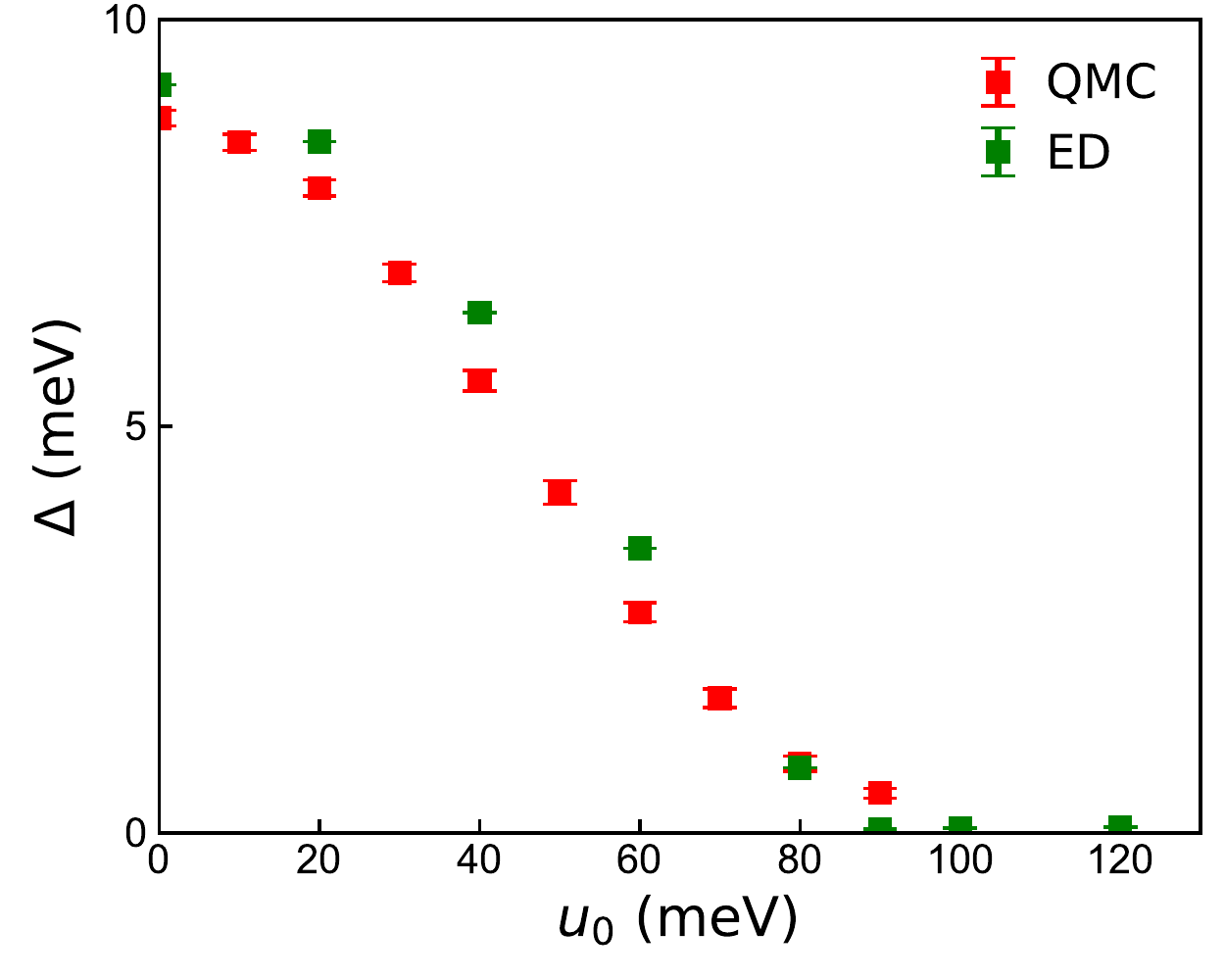}
	\caption{Quasiparticle gaps of $\Gamma$ from ED (green squares) when $L=3$ compared with counterparts from QMC (red squres) with $T=3$ meV.}
	\label{fig:GapwithED}
\end{figure}

The gaps of $L=3$ from ED, considering the computational complexity, is calculated as well and shown in Fig.~\ref{fig:GapwithED} (green squares), where counterparts from QMC (red squares), with the simulation temperature being 3 meV, are also shown. As the result from ED is of ground state, gaps from QMC at 3 meV are slightly lower than those from ED, and the lowering of simulation $T$ can reduce the deviation. Besides, the same tendency is obtained that the gap closes at $u_0\approx0.8u_1$. In addition, the detail of implementaion of ED is depicted in Appendix G.

\section{Appendix E: $C_3$ rotation symmetry of the semimetallic phase}

\begin{figure}[!h]
	\centering
	\includegraphics[width=0.7\linewidth]{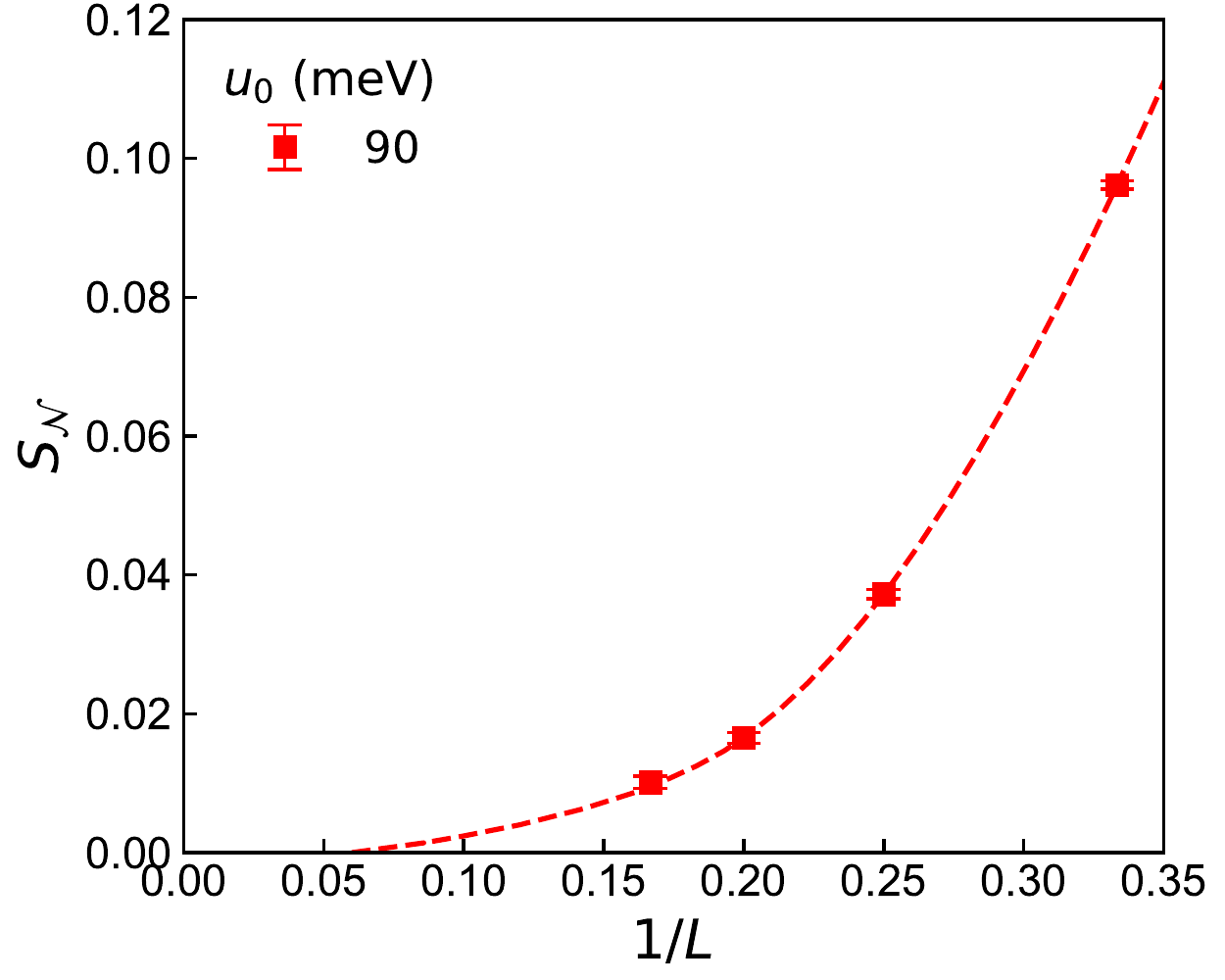}
	\caption{Correlation, $S_\mathcal{N}$, of the nematic order parameter, $\mathcal{N}$, as a function of reversed linear size, $1/L$.}
	\label{fig:S_C}
\end{figure}

The correlation of the nematic order, $\mathcal{N}$, as a function of reversed linear size, $1/L$, with $u_0=$ 90 meV is shown in Fig. \ref{fig:S_C}. As there might be some correlation in a small system, there is none in the thermodynamic limit, suggesting the absence of this nematic order and the presence of $C_3$ rotation symmetry.

\section{Appendix F: Implementation of the exact diagonalization}
Exact diagonalization (ED) is implemented to obtain the charge gap of the system in comparison to those from QMC. For $L\times L$ ($\equiv N_k$) systems with the two nearly flat bands taken into account, there are $2L^2$ single particle states in total which are labeled by momentum and band index. Here we use an integer $i\in [0,2L^2-1]$ to label each single particle state, $d^\dagger_i\equiv d^\dagger_{\mathbf km}$, whose momentum $\mathbf k$ and band index $m$ are determined by $i$. The Hamiltonian in Eq. (6) can be rewritten in the following form by moving all electron creation operator to the left:
\begin{equation}
    H=\sum_{i,i',j',j} V_2(i,i',j',j) d_i^\dagger d_{i'}^\dagger d_{j'}d_j+\sum_{i,j}V_1(i,j)d_i^\dagger  d_j+V_0.
    \label{EDsim}
\end{equation}
The matrix element between arbitrary manybody states can be obtained from Eq. \eqref{EDsim}. Let $\ket{\psi}$ be a simple direct product state $\ket{\psi}=C^\dagger_{i_{N-1}}...C^\dagger_{i_0}\ket{0}$ whose creation operators are ordered by $i_{N-1}>...>i_1>i_0$. Then $\ket{\psi}$ can also be represented by a string of bits with length $2L^2$ such as $\ket{1101001...}$. The Hamiltonian can be written in the basis made of these distinct direct product states. For a given product state $|\psi\rangle$, if there is another product state $|\psi'\rangle$ such that $\langle\psi'|H|\psi\rangle\ne 0$, then either of the two situations occurs: (1) there exist distinct numbers $i\ne i'$ such that $|\psi_1'\rangle\propto d^\dagger_{i'} d_i|\psi\rangle$; (2) there exist distinct numbers $i,j,i',j'$ with $i>j,i'>j'$ such that $|\psi_2'\rangle\propto d^\dagger_{j'}d^\dagger_{i'} d_i d_j|\psi\rangle$. Then, according to Eq.\eqref{EDsim}, we have
\begin{equation}
\begin{aligned}
\bra{\psi_2'} H\ket{\psi}=&(-1)^{\mu(i,j)}(-1)^{\mu'(i',j')}[V_2(j',i',i,j)\\
&  -V_2(j',i',j,i)-V_2(i',j',i,j)+V_2(i',j',j,i)].
\label{ED2}
\end{aligned}
\end{equation}
Here $\mu(i,j)$ is the number of "1" between the $i$-th and $j$-th bits in the bit-representation of $\ket{\psi}$. This sign factor comes from the Fermi anti-commutation of creation operators. Similarly, for $|\psi_1'\rangle\propto d^\dagger_{i'} d_i|\psi\rangle$, we have
\begin{equation}
\begin{aligned}
\bra{\psi_1'} H\ket{\psi}=\sum_{j}&(-1)^{\mu(i,j)}(-1)^{\mu'(i',j)}\left[V_2(j,i',i,j)\right.\\
-&\left.V_2(i',j,i,j)-V_2(j,i',j,i)+V_2(i',j,j,i)\right] \\
&+V_1(i',i).
\label{ED1}
\end{aligned}
\end{equation}
The diagonal matrix elements are
\begin{equation}
\begin{aligned}
\bra{\psi}H\ket{\psi}=&V_0+\sum_{i>j,\psi_i=\psi_j="1"}[V_2(j,i,i,j)-V_2(j,i,j,i) \\
&-V_2(i,j,i,j)+V_2(i,j,j,i)]+\sum_{i,\psi_i="1"}V_1(i,i).
\label{ED0}
\end{aligned}
\end{equation}
Then $H$ can be diagonalized in the basis of $|\psi\rangle$ to obtain its eigenvalues.

\bibliography{ref}

\begin{thebibliography}{70}%
\makeatletter
\providecommand \@ifxundefined [1]{%
 \@ifx{#1\undefined}
}%
\providecommand \@ifnum [1]{%
 \ifnum #1\expandafter \@firstoftwo
 \else \expandafter \@secondoftwo
 \fi
}%
\providecommand \@ifx [1]{%
 \ifx #1\expandafter \@firstoftwo
 \else \expandafter \@secondoftwo
 \fi
}%
\providecommand \natexlab [1]{#1}%
\providecommand \enquote  [1]{``#1''}%
\providecommand \bibnamefont  [1]{#1}%
\providecommand \bibfnamefont [1]{#1}%
\providecommand \citenamefont [1]{#1}%
\providecommand \href@noop [0]{\@secondoftwo}%
\providecommand \href [0]{\begingroup \@sanitize@url \@href}%
\providecommand \@href[1]{\@@startlink{#1}\@@href}%
\providecommand \@@href[1]{\endgroup#1\@@endlink}%
\providecommand \@sanitize@url [0]{\catcode `\\12\catcode `\$12\catcode
  `\&12\catcode `\#12\catcode `\^12\catcode `\_12\catcode `\%12\relax}%
\providecommand \@@startlink[1]{}%
\providecommand \@@endlink[0]{}%
\providecommand \url  [0]{\begingroup\@sanitize@url \@url }%
\providecommand \@url [1]{\endgroup\@href {#1}{\urlprefix }}%
\providecommand \urlprefix  [0]{URL }%
\providecommand \Eprint [0]{\href }%
\providecommand \doibase [0]{https://doi.org/}%
\providecommand \selectlanguage [0]{\@gobble}%
\providecommand \bibinfo  [0]{\@secondoftwo}%
\providecommand \bibfield  [0]{\@secondoftwo}%
\providecommand \translation [1]{[#1]}%
\providecommand \BibitemOpen [0]{}%
\providecommand \bibitemStop [0]{}%
\providecommand \bibitemNoStop [0]{.\EOS\space}%
\providecommand \EOS [0]{\spacefactor3000\relax}%
\providecommand \BibitemShut  [1]{\csname bibitem#1\endcsname}%
\let\auto@bib@innerbib\@empty
\bibitem [{\citenamefont {Trambly~de Laissardière}\ \emph
  {et~al.}(2010)\citenamefont {Trambly~de Laissardière}, \citenamefont
  {Mayou},\ and\ \citenamefont {Magaud}}]{tramblyLocalization2010}%
  \BibitemOpen
  \bibfield  {author} {\bibinfo {author} {\bibfnamefont {G.}~\bibnamefont
  {Trambly~de Laissardière}}, \bibinfo {author} {\bibfnamefont
  {D.}~\bibnamefont {Mayou}},\ and\ \bibinfo {author} {\bibfnamefont
  {L.}~\bibnamefont {Magaud}},\ }\bibfield  {title} {\bibinfo {title}
  {Localization of dirac electrons in rotated graphene bilayers},\ }\href
  {https://doi.org/doi: 10.1021/nl902948m} {\bibfield  {journal} {\bibinfo
  {journal} {Nano Letters}\ }\textbf {\bibinfo {volume} {10}},\ \bibinfo
  {pages} {804 } (\bibinfo {year} {2010})}\BibitemShut {NoStop}%
\bibitem [{\citenamefont {Trambly~de Laissardi\`ere}\ \emph
  {et~al.}(2012)\citenamefont {Trambly~de Laissardi\`ere}, \citenamefont
  {Mayou},\ and\ \citenamefont {Magaud}}]{tramblyNumerical2012}%
  \BibitemOpen
  \bibfield  {author} {\bibinfo {author} {\bibfnamefont {G.}~\bibnamefont
  {Trambly~de Laissardi\`ere}}, \bibinfo {author} {\bibfnamefont
  {D.}~\bibnamefont {Mayou}},\ and\ \bibinfo {author} {\bibfnamefont
  {L.}~\bibnamefont {Magaud}},\ }\bibfield  {title} {\bibinfo {title}
  {Numerical studies of confined states in rotated bilayers of graphene},\
  }\href {https://doi.org/10.1103/PhysRevB.86.125413} {\bibfield  {journal}
  {\bibinfo  {journal} {Phys. Rev. B}\ }\textbf {\bibinfo {volume} {86}},\
  \bibinfo {pages} {125413} (\bibinfo {year} {2012})}\BibitemShut {NoStop}%
\bibitem [{\citenamefont {Bistritzer}\ and\ \citenamefont
  {MacDonald}(2011)}]{rafiMoire2011}%
  \BibitemOpen
  \bibfield  {author} {\bibinfo {author} {\bibfnamefont {R.}~\bibnamefont
  {Bistritzer}}\ and\ \bibinfo {author} {\bibfnamefont {A.~H.}\ \bibnamefont
  {MacDonald}},\ }\bibfield  {title} {\bibinfo {title} {Moir\'e bands in
  twisted double-layer graphene},\ }\href
  {https://doi.org/10.1073/pnas.1108174108} {\bibfield  {journal} {\bibinfo
  {journal} {Proceedings of the National Academy of Sciences}\ }\textbf
  {\bibinfo {volume} {108}},\ \bibinfo {pages} {12233} (\bibinfo {year}
  {2011})}\BibitemShut {NoStop}%
\bibitem [{\citenamefont {Cao}\ \emph {et~al.}(2018{\natexlab{a}})\citenamefont
  {Cao}, \citenamefont {Fatemi}, \citenamefont {Demir}, \citenamefont {Fang},
  \citenamefont {Tomarken}, \citenamefont {Luo}, \citenamefont
  {Sanchez-Yamagishi}, \citenamefont {Watanabe}, \citenamefont {Taniguchi},
  \citenamefont {Kaxiras} \emph {et~al.}}]{caoCorrelated2018}%
  \BibitemOpen
  \bibfield  {author} {\bibinfo {author} {\bibfnamefont {Y.}~\bibnamefont
  {Cao}}, \bibinfo {author} {\bibfnamefont {V.}~\bibnamefont {Fatemi}},
  \bibinfo {author} {\bibfnamefont {A.}~\bibnamefont {Demir}}, \bibinfo
  {author} {\bibfnamefont {S.}~\bibnamefont {Fang}}, \bibinfo {author}
  {\bibfnamefont {S.~L.}\ \bibnamefont {Tomarken}}, \bibinfo {author}
  {\bibfnamefont {J.~Y.}\ \bibnamefont {Luo}}, \bibinfo {author} {\bibfnamefont
  {J.~D.}\ \bibnamefont {Sanchez-Yamagishi}}, \bibinfo {author} {\bibfnamefont
  {K.}~\bibnamefont {Watanabe}}, \bibinfo {author} {\bibfnamefont
  {T.}~\bibnamefont {Taniguchi}}, \bibinfo {author} {\bibfnamefont
  {E.}~\bibnamefont {Kaxiras}}, \emph {et~al.},\ }\bibfield  {title} {\bibinfo
  {title} {Correlated insulator behaviour at half-filling in magic-angle
  graphene superlattices},\ }\href {https://doi.org/10.1038/nature26154}
  {\bibfield  {journal} {\bibinfo  {journal} {Nature}\ }\textbf {\bibinfo
  {volume} {556}},\ \bibinfo {pages} {80} (\bibinfo {year}
  {2018}{\natexlab{a}})}\BibitemShut {NoStop}%
\bibitem [{\citenamefont {Cao}\ \emph {et~al.}(2020)\citenamefont {Cao},
  \citenamefont {Rodan-Legrain}, \citenamefont {Rubies-Bigorda}, \citenamefont
  {Park}, \citenamefont {Watanabe}, \citenamefont {Taniguchi},\ and\
  \citenamefont {Jarillo-Herrero}}]{caoTunable2020}%
  \BibitemOpen
  \bibfield  {author} {\bibinfo {author} {\bibfnamefont {Y.}~\bibnamefont
  {Cao}}, \bibinfo {author} {\bibfnamefont {D.}~\bibnamefont {Rodan-Legrain}},
  \bibinfo {author} {\bibfnamefont {O.}~\bibnamefont {Rubies-Bigorda}},
  \bibinfo {author} {\bibfnamefont {J.~M.}\ \bibnamefont {Park}}, \bibinfo
  {author} {\bibfnamefont {K.}~\bibnamefont {Watanabe}}, \bibinfo {author}
  {\bibfnamefont {T.}~\bibnamefont {Taniguchi}},\ and\ \bibinfo {author}
  {\bibfnamefont {P.}~\bibnamefont {Jarillo-Herrero}},\ }\bibfield  {title}
  {\bibinfo {title} {Tunable correlated states and spin-polarized phases in
  twisted bilayer--bilayer graphene},\ }\href
  {https://doi.org/10.1038/s41586-020-2260-6} {\bibfield  {journal} {\bibinfo
  {journal} {Nature}\ }\textbf {\bibinfo {volume} {583}},\ \bibinfo {pages}
  {215} (\bibinfo {year} {2020})}\BibitemShut {NoStop}%
\bibitem [{\citenamefont {Polshyn}\ \emph {et~al.}(2019)\citenamefont
  {Polshyn}, \citenamefont {Yankowitz}, \citenamefont {Chen}, \citenamefont
  {Zhang}, \citenamefont {Watanabe}, \citenamefont {Taniguchi}, \citenamefont
  {Dean},\ and\ \citenamefont {Young}}]{polshynLarge2019}%
  \BibitemOpen
  \bibfield  {author} {\bibinfo {author} {\bibfnamefont {H.}~\bibnamefont
  {Polshyn}}, \bibinfo {author} {\bibfnamefont {M.}~\bibnamefont {Yankowitz}},
  \bibinfo {author} {\bibfnamefont {S.}~\bibnamefont {Chen}}, \bibinfo {author}
  {\bibfnamefont {Y.}~\bibnamefont {Zhang}}, \bibinfo {author} {\bibfnamefont
  {K.}~\bibnamefont {Watanabe}}, \bibinfo {author} {\bibfnamefont
  {T.}~\bibnamefont {Taniguchi}}, \bibinfo {author} {\bibfnamefont {C.~R.}\
  \bibnamefont {Dean}},\ and\ \bibinfo {author} {\bibfnamefont {A.~F.}\
  \bibnamefont {Young}},\ }\bibfield  {title} {\bibinfo {title} {Large
  linear-in-temperature resistivity in twisted bilayer graphene},\ }\href
  {https://doi.org/10.1038/s41567-019-0596-3} {\bibfield  {journal} {\bibinfo
  {journal} {Nature Physics}\ }\textbf {\bibinfo {volume} {15}},\ \bibinfo
  {pages} {1011} (\bibinfo {year} {2019})}\BibitemShut {NoStop}%
\bibitem [{\citenamefont {Liu}\ \emph {et~al.}(2021{\natexlab{a}})\citenamefont
  {Liu}, \citenamefont {Wang}, \citenamefont {Watanabe}, \citenamefont
  {Taniguchi}, \citenamefont {Vafek},\ and\ \citenamefont
  {Li}}]{liuTuning2021}%
  \BibitemOpen
  \bibfield  {author} {\bibinfo {author} {\bibfnamefont {X.}~\bibnamefont
  {Liu}}, \bibinfo {author} {\bibfnamefont {Z.}~\bibnamefont {Wang}}, \bibinfo
  {author} {\bibfnamefont {K.}~\bibnamefont {Watanabe}}, \bibinfo {author}
  {\bibfnamefont {T.}~\bibnamefont {Taniguchi}}, \bibinfo {author}
  {\bibfnamefont {O.}~\bibnamefont {Vafek}},\ and\ \bibinfo {author}
  {\bibfnamefont {J.}~\bibnamefont {Li}},\ }\bibfield  {title} {\bibinfo
  {title} {Tuning electron correlation in magic-angle twisted bilayer graphene
  using coulomb screening},\ }\href {https://doi.org/DOI:
  10.1126/science.abb8754} {\bibfield  {journal} {\bibinfo  {journal}
  {Science}\ }\textbf {\bibinfo {volume} {371}},\ \bibinfo {pages} {1261}
  (\bibinfo {year} {2021}{\natexlab{a}})}\BibitemShut {NoStop}%
\bibitem [{\citenamefont {Xie}\ \emph {et~al.}(2019)\citenamefont {Xie},
  \citenamefont {Lian}, \citenamefont {J{\"a}ck}, \citenamefont {Liu},
  \citenamefont {Chiu}, \citenamefont {Watanabe}, \citenamefont {Taniguchi},
  \citenamefont {Bernevig},\ and\ \citenamefont
  {Yazdani}}]{xie2019spectroscopic}%
  \BibitemOpen
  \bibfield  {author} {\bibinfo {author} {\bibfnamefont {Y.}~\bibnamefont
  {Xie}}, \bibinfo {author} {\bibfnamefont {B.}~\bibnamefont {Lian}}, \bibinfo
  {author} {\bibfnamefont {B.}~\bibnamefont {J{\"a}ck}}, \bibinfo {author}
  {\bibfnamefont {X.}~\bibnamefont {Liu}}, \bibinfo {author} {\bibfnamefont
  {C.-L.}\ \bibnamefont {Chiu}}, \bibinfo {author} {\bibfnamefont
  {K.}~\bibnamefont {Watanabe}}, \bibinfo {author} {\bibfnamefont
  {T.}~\bibnamefont {Taniguchi}}, \bibinfo {author} {\bibfnamefont {B.~A.}\
  \bibnamefont {Bernevig}},\ and\ \bibinfo {author} {\bibfnamefont
  {A.}~\bibnamefont {Yazdani}},\ }\bibfield  {title} {\bibinfo {title}
  {Spectroscopic signatures of many-body correlations in magic-angle twisted
  bilayer graphene},\ }\href {https://doi.org/10.1038/s41586-019-1422-x}
  {\bibfield  {journal} {\bibinfo  {journal} {Nature}\ }\textbf {\bibinfo
  {volume} {572}},\ \bibinfo {pages} {101} (\bibinfo {year}
  {2019})}\BibitemShut {NoStop}%
\bibitem [{\citenamefont {Choi}\ \emph {et~al.}(2019)\citenamefont {Choi},
  \citenamefont {Kemmer}, \citenamefont {Peng}, \citenamefont {Thomson},
  \citenamefont {Arora}, \citenamefont {Polski}, \citenamefont {Zhang},
  \citenamefont {Ren}, \citenamefont {Alicea}, \citenamefont {Refael} \emph
  {et~al.}}]{choi2019electronic}%
  \BibitemOpen
  \bibfield  {author} {\bibinfo {author} {\bibfnamefont {Y.}~\bibnamefont
  {Choi}}, \bibinfo {author} {\bibfnamefont {J.}~\bibnamefont {Kemmer}},
  \bibinfo {author} {\bibfnamefont {Y.}~\bibnamefont {Peng}}, \bibinfo {author}
  {\bibfnamefont {A.}~\bibnamefont {Thomson}}, \bibinfo {author} {\bibfnamefont
  {H.}~\bibnamefont {Arora}}, \bibinfo {author} {\bibfnamefont
  {R.}~\bibnamefont {Polski}}, \bibinfo {author} {\bibfnamefont
  {Y.}~\bibnamefont {Zhang}}, \bibinfo {author} {\bibfnamefont
  {H.}~\bibnamefont {Ren}}, \bibinfo {author} {\bibfnamefont {J.}~\bibnamefont
  {Alicea}}, \bibinfo {author} {\bibfnamefont {G.}~\bibnamefont {Refael}},
  \emph {et~al.},\ }\bibfield  {title} {\bibinfo {title} {Electronic
  correlations in twisted bilayer graphene near the magic angle},\ }\href
  {https://doi.org/10.1038/s41567-019-0606-5} {\bibfield  {journal} {\bibinfo
  {journal} {Nature physics}\ }\textbf {\bibinfo {volume} {15}},\ \bibinfo
  {pages} {1174} (\bibinfo {year} {2019})}\BibitemShut {NoStop}%
\bibitem [{\citenamefont {Nuckolls}\ \emph {et~al.}(2020)\citenamefont
  {Nuckolls}, \citenamefont {Oh}, \citenamefont {Wong}, \citenamefont {Lian},
  \citenamefont {Watanabe}, \citenamefont {Taniguchi}, \citenamefont
  {Bernevig},\ and\ \citenamefont {Yazdani}}]{nuckolls2020strongly}%
  \BibitemOpen
  \bibfield  {author} {\bibinfo {author} {\bibfnamefont {K.~P.}\ \bibnamefont
  {Nuckolls}}, \bibinfo {author} {\bibfnamefont {M.}~\bibnamefont {Oh}},
  \bibinfo {author} {\bibfnamefont {D.}~\bibnamefont {Wong}}, \bibinfo {author}
  {\bibfnamefont {B.}~\bibnamefont {Lian}}, \bibinfo {author} {\bibfnamefont
  {K.}~\bibnamefont {Watanabe}}, \bibinfo {author} {\bibfnamefont
  {T.}~\bibnamefont {Taniguchi}}, \bibinfo {author} {\bibfnamefont {B.~A.}\
  \bibnamefont {Bernevig}},\ and\ \bibinfo {author} {\bibfnamefont
  {A.}~\bibnamefont {Yazdani}},\ }\bibfield  {title} {\bibinfo {title}
  {Strongly correlated chern insulators in magic-angle twisted bilayer
  graphene},\ }\href {https://doi.org/10.1038/s41586-020-3028-8} {\bibfield
  {journal} {\bibinfo  {journal} {Nature}\ }\textbf {\bibinfo {volume} {588}},\
  \bibinfo {pages} {610} (\bibinfo {year} {2020})}\BibitemShut {NoStop}%
\bibitem [{\citenamefont {Saito}\ \emph {et~al.}(2021)\citenamefont {Saito},
  \citenamefont {Ge}, \citenamefont {Rademaker}, \citenamefont {Watanabe},
  \citenamefont {Taniguchi}, \citenamefont {Abanin},\ and\ \citenamefont
  {Young}}]{saito2021hofstadter}%
  \BibitemOpen
  \bibfield  {author} {\bibinfo {author} {\bibfnamefont {Y.}~\bibnamefont
  {Saito}}, \bibinfo {author} {\bibfnamefont {J.}~\bibnamefont {Ge}}, \bibinfo
  {author} {\bibfnamefont {L.}~\bibnamefont {Rademaker}}, \bibinfo {author}
  {\bibfnamefont {K.}~\bibnamefont {Watanabe}}, \bibinfo {author}
  {\bibfnamefont {T.}~\bibnamefont {Taniguchi}}, \bibinfo {author}
  {\bibfnamefont {D.~A.}\ \bibnamefont {Abanin}},\ and\ \bibinfo {author}
  {\bibfnamefont {A.~F.}\ \bibnamefont {Young}},\ }\bibfield  {title} {\bibinfo
  {title} {Hofstadter subband ferromagnetism and symmetry-broken chern
  insulators in twisted bilayer graphene},\ }\href
  {https://doi.org/10.1038/s41567-020-01129-4} {\bibfield  {journal} {\bibinfo
  {journal} {Nature Physics}\ }\textbf {\bibinfo {volume} {17}},\ \bibinfo
  {pages} {478} (\bibinfo {year} {2021})}\BibitemShut {NoStop}%
\bibitem [{\citenamefont {Das}\ \emph {et~al.}(2021)\citenamefont {Das},
  \citenamefont {Lu}, \citenamefont {Herzog-Arbeitman}, \citenamefont {Song},
  \citenamefont {Watanabe}, \citenamefont {Taniguchi}, \citenamefont
  {Bernevig},\ and\ \citenamefont {Efetov}}]{das2021symmetry}%
  \BibitemOpen
  \bibfield  {author} {\bibinfo {author} {\bibfnamefont {I.}~\bibnamefont
  {Das}}, \bibinfo {author} {\bibfnamefont {X.}~\bibnamefont {Lu}}, \bibinfo
  {author} {\bibfnamefont {J.}~\bibnamefont {Herzog-Arbeitman}}, \bibinfo
  {author} {\bibfnamefont {Z.-D.}\ \bibnamefont {Song}}, \bibinfo {author}
  {\bibfnamefont {K.}~\bibnamefont {Watanabe}}, \bibinfo {author}
  {\bibfnamefont {T.}~\bibnamefont {Taniguchi}}, \bibinfo {author}
  {\bibfnamefont {B.~A.}\ \bibnamefont {Bernevig}},\ and\ \bibinfo {author}
  {\bibfnamefont {D.~K.}\ \bibnamefont {Efetov}},\ }\bibfield  {title}
  {\bibinfo {title} {Symmetry-broken chern insulators and rashba-like
  landau-level crossings in magic-angle bilayer graphene},\ }\href
  {https://doi.org/10.1038/s41567-021-01186-3} {\bibfield  {journal} {\bibinfo
  {journal} {Nature Physics}\ }\textbf {\bibinfo {volume} {17}},\ \bibinfo
  {pages} {710} (\bibinfo {year} {2021})}\BibitemShut {NoStop}%
\bibitem [{\citenamefont {Wu}\ \emph {et~al.}(2021)\citenamefont {Wu},
  \citenamefont {Zhang}, \citenamefont {Watanabe}, \citenamefont {Taniguchi},\
  and\ \citenamefont {Andrei}}]{wu2021chern}%
  \BibitemOpen
  \bibfield  {author} {\bibinfo {author} {\bibfnamefont {S.}~\bibnamefont
  {Wu}}, \bibinfo {author} {\bibfnamefont {Z.}~\bibnamefont {Zhang}}, \bibinfo
  {author} {\bibfnamefont {K.}~\bibnamefont {Watanabe}}, \bibinfo {author}
  {\bibfnamefont {T.}~\bibnamefont {Taniguchi}},\ and\ \bibinfo {author}
  {\bibfnamefont {E.~Y.}\ \bibnamefont {Andrei}},\ }\bibfield  {title}
  {\bibinfo {title} {Chern insulators, van hove singularities and topological
  flat bands in magic-angle twisted bilayer graphene},\ }\href
  {https://doi.org/10.1038/s41563-020-00911-2} {\bibfield  {journal} {\bibinfo
  {journal} {Nature materials}\ }\textbf {\bibinfo {volume} {20}},\ \bibinfo
  {pages} {488} (\bibinfo {year} {2021})}\BibitemShut {NoStop}%
\bibitem [{\citenamefont {Potasz}\ \emph {et~al.}(2021)\citenamefont {Potasz},
  \citenamefont {Xie},\ and\ \citenamefont {MacDonald}}]{potaszExact2021}%
  \BibitemOpen
  \bibfield  {author} {\bibinfo {author} {\bibfnamefont {P.}~\bibnamefont
  {Potasz}}, \bibinfo {author} {\bibfnamefont {M.}~\bibnamefont {Xie}},\ and\
  \bibinfo {author} {\bibfnamefont {A.~H.}\ \bibnamefont {MacDonald}},\
  }\bibfield  {title} {\bibinfo {title} {Exact diagonalization for magic-angle
  twisted bilayer graphene},\ }\href
  {https://doi.org/10.1103/PhysRevLett.127.147203} {\bibfield  {journal}
  {\bibinfo  {journal} {Phys. Rev. Lett.}\ }\textbf {\bibinfo {volume} {127}},\
  \bibinfo {pages} {147203} (\bibinfo {year} {2021})}\BibitemShut {NoStop}%
\bibitem [{\citenamefont {Jaoui}\ \emph {et~al.}(2022)\citenamefont {Jaoui},
  \citenamefont {Das}, \citenamefont {Di~Battista}, \citenamefont
  {D{\'\i}ez-M{\'e}rida}, \citenamefont {Lu}, \citenamefont {Watanabe},
  \citenamefont {Taniguchi}, \citenamefont {Ishizuka}, \citenamefont
  {Levitov},\ and\ \citenamefont {Efetov}}]{jaoui2022quantum}%
  \BibitemOpen
  \bibfield  {author} {\bibinfo {author} {\bibfnamefont {A.}~\bibnamefont
  {Jaoui}}, \bibinfo {author} {\bibfnamefont {I.}~\bibnamefont {Das}}, \bibinfo
  {author} {\bibfnamefont {G.}~\bibnamefont {Di~Battista}}, \bibinfo {author}
  {\bibfnamefont {J.}~\bibnamefont {D{\'\i}ez-M{\'e}rida}}, \bibinfo {author}
  {\bibfnamefont {X.}~\bibnamefont {Lu}}, \bibinfo {author} {\bibfnamefont
  {K.}~\bibnamefont {Watanabe}}, \bibinfo {author} {\bibfnamefont
  {T.}~\bibnamefont {Taniguchi}}, \bibinfo {author} {\bibfnamefont
  {H.}~\bibnamefont {Ishizuka}}, \bibinfo {author} {\bibfnamefont
  {L.}~\bibnamefont {Levitov}},\ and\ \bibinfo {author} {\bibfnamefont {D.~K.}\
  \bibnamefont {Efetov}},\ }\bibfield  {title} {\bibinfo {title} {Quantum
  critical behaviour in magic-angle twisted bilayer graphene},\ }\href
  {https://doi.org/10.1038/s41567-022-01556-5} {\bibfield  {journal} {\bibinfo
  {journal} {Nature Physics}\ }\textbf {\bibinfo {volume} {18}},\ \bibinfo
  {pages} {633} (\bibinfo {year} {2022})}\BibitemShut {NoStop}%
\bibitem [{\citenamefont {Cao}\ \emph {et~al.}(2018{\natexlab{b}})\citenamefont
  {Cao}, \citenamefont {Fatemi}, \citenamefont {Fang}, \citenamefont
  {Watanabe}, \citenamefont {Taniguchi}, \citenamefont {Kaxiras},\ and\
  \citenamefont {Jarillo-Herrero}}]{caoUnconventional2018}%
  \BibitemOpen
  \bibfield  {author} {\bibinfo {author} {\bibfnamefont {Y.}~\bibnamefont
  {Cao}}, \bibinfo {author} {\bibfnamefont {V.}~\bibnamefont {Fatemi}},
  \bibinfo {author} {\bibfnamefont {S.}~\bibnamefont {Fang}}, \bibinfo {author}
  {\bibfnamefont {K.}~\bibnamefont {Watanabe}}, \bibinfo {author}
  {\bibfnamefont {T.}~\bibnamefont {Taniguchi}}, \bibinfo {author}
  {\bibfnamefont {E.}~\bibnamefont {Kaxiras}},\ and\ \bibinfo {author}
  {\bibfnamefont {P.}~\bibnamefont {Jarillo-Herrero}},\ }\bibfield  {title}
  {\bibinfo {title} {Unconventional superconductivity in magic-angle graphene
  superlattices},\ }\href {https://doi.org/10.1038/nature26160} {\bibfield
  {journal} {\bibinfo  {journal} {Nature}\ }\textbf {\bibinfo {volume} {556}},\
  \bibinfo {pages} {43} (\bibinfo {year} {2018}{\natexlab{b}})}\BibitemShut
  {NoStop}%
\bibitem [{\citenamefont {Cao}\ \emph {et~al.}(2021)\citenamefont {Cao},
  \citenamefont {Rodan-Legrain}, \citenamefont {Park}, \citenamefont {Yuan},
  \citenamefont {Watanabe}, \citenamefont {Taniguchi}, \citenamefont
  {Fernandes}, \citenamefont {Fu},\ and\ \citenamefont
  {Jarillo-Herrero}}]{cao2021nematicity}%
  \BibitemOpen
  \bibfield  {author} {\bibinfo {author} {\bibfnamefont {Y.}~\bibnamefont
  {Cao}}, \bibinfo {author} {\bibfnamefont {D.}~\bibnamefont {Rodan-Legrain}},
  \bibinfo {author} {\bibfnamefont {J.~M.}\ \bibnamefont {Park}}, \bibinfo
  {author} {\bibfnamefont {N.~F.}\ \bibnamefont {Yuan}}, \bibinfo {author}
  {\bibfnamefont {K.}~\bibnamefont {Watanabe}}, \bibinfo {author}
  {\bibfnamefont {T.}~\bibnamefont {Taniguchi}}, \bibinfo {author}
  {\bibfnamefont {R.~M.}\ \bibnamefont {Fernandes}}, \bibinfo {author}
  {\bibfnamefont {L.}~\bibnamefont {Fu}},\ and\ \bibinfo {author}
  {\bibfnamefont {P.}~\bibnamefont {Jarillo-Herrero}},\ }\bibfield  {title}
  {\bibinfo {title} {Nematicity and competing orders in superconducting
  magic-angle graphene},\ }\href {https://doi.org/10.1126/science.abc2836}
  {\bibfield  {journal} {\bibinfo  {journal} {science}\ }\textbf {\bibinfo
  {volume} {372}},\ \bibinfo {pages} {264} (\bibinfo {year}
  {2021})}\BibitemShut {NoStop}%
\bibitem [{\citenamefont {Yankowitz}\ \emph {et~al.}(2019)\citenamefont
  {Yankowitz}, \citenamefont {Chen}, \citenamefont {Polshyn}, \citenamefont
  {Zhang}, \citenamefont {Watanabe}, \citenamefont {Taniguchi}, \citenamefont
  {Graf}, \citenamefont {Young},\ and\ \citenamefont
  {Dean}}]{yankowitz2019tuning}%
  \BibitemOpen
  \bibfield  {author} {\bibinfo {author} {\bibfnamefont {M.}~\bibnamefont
  {Yankowitz}}, \bibinfo {author} {\bibfnamefont {S.}~\bibnamefont {Chen}},
  \bibinfo {author} {\bibfnamefont {H.}~\bibnamefont {Polshyn}}, \bibinfo
  {author} {\bibfnamefont {Y.}~\bibnamefont {Zhang}}, \bibinfo {author}
  {\bibfnamefont {K.}~\bibnamefont {Watanabe}}, \bibinfo {author}
  {\bibfnamefont {T.}~\bibnamefont {Taniguchi}}, \bibinfo {author}
  {\bibfnamefont {D.}~\bibnamefont {Graf}}, \bibinfo {author} {\bibfnamefont
  {A.~F.}\ \bibnamefont {Young}},\ and\ \bibinfo {author} {\bibfnamefont
  {C.~R.}\ \bibnamefont {Dean}},\ }\bibfield  {title} {\bibinfo {title} {Tuning
  superconductivity in twisted bilayer graphene},\ }\href {https://doi.org/DOI:
  10.1126/science.aav1910} {\bibfield  {journal} {\bibinfo  {journal}
  {Science}\ }\textbf {\bibinfo {volume} {363}},\ \bibinfo {pages} {1059}
  (\bibinfo {year} {2019})}\BibitemShut {NoStop}%
\bibitem [{\citenamefont {Diez-Merida}\ \emph {et~al.}(2021)\citenamefont
  {Diez-Merida}, \citenamefont {D{\'\i}ez-Carl{\'o}n}, \citenamefont {Yang},
  \citenamefont {Xie}, \citenamefont {Gao}, \citenamefont {Watanabe},
  \citenamefont {Taniguchi}, \citenamefont {Lu}, \citenamefont {Law},\ and\
  \citenamefont {Efetov}}]{diez2021magnetic}%
  \BibitemOpen
  \bibfield  {author} {\bibinfo {author} {\bibfnamefont {J.}~\bibnamefont
  {Diez-Merida}}, \bibinfo {author} {\bibfnamefont {A.}~\bibnamefont
  {D{\'\i}ez-Carl{\'o}n}}, \bibinfo {author} {\bibfnamefont {S.}~\bibnamefont
  {Yang}}, \bibinfo {author} {\bibfnamefont {Y.-M.}\ \bibnamefont {Xie}},
  \bibinfo {author} {\bibfnamefont {X.-J.}\ \bibnamefont {Gao}}, \bibinfo
  {author} {\bibfnamefont {K.}~\bibnamefont {Watanabe}}, \bibinfo {author}
  {\bibfnamefont {T.}~\bibnamefont {Taniguchi}}, \bibinfo {author}
  {\bibfnamefont {X.}~\bibnamefont {Lu}}, \bibinfo {author} {\bibfnamefont
  {K.~T.}\ \bibnamefont {Law}},\ and\ \bibinfo {author} {\bibfnamefont {D.~K.}\
  \bibnamefont {Efetov}},\ }\bibfield  {title} {\bibinfo {title} {Magnetic
  josephson junctions and superconducting diodes in magic angle twisted bilayer
  graphene},\ }\href {https://doi.org/10.48550/arXiv.2110.01067} {\bibfield
  {journal} {\bibinfo  {journal} {arXiv preprint arXiv:2110.01067}\ } (\bibinfo
  {year} {2021})}\BibitemShut {NoStop}%
\bibitem [{\citenamefont {Di~Battista}\ \emph {et~al.}(2022)\citenamefont
  {Di~Battista}, \citenamefont {Seifert}, \citenamefont {Watanabe},
  \citenamefont {Taniguchi}, \citenamefont {Fong}, \citenamefont {Principi},\
  and\ \citenamefont {Efetov}}]{di2022revealing}%
  \BibitemOpen
  \bibfield  {author} {\bibinfo {author} {\bibfnamefont {G.}~\bibnamefont
  {Di~Battista}}, \bibinfo {author} {\bibfnamefont {P.}~\bibnamefont
  {Seifert}}, \bibinfo {author} {\bibfnamefont {K.}~\bibnamefont {Watanabe}},
  \bibinfo {author} {\bibfnamefont {T.}~\bibnamefont {Taniguchi}}, \bibinfo
  {author} {\bibfnamefont {K.~C.}\ \bibnamefont {Fong}}, \bibinfo {author}
  {\bibfnamefont {A.}~\bibnamefont {Principi}},\ and\ \bibinfo {author}
  {\bibfnamefont {D.~K.}\ \bibnamefont {Efetov}},\ }\bibfield  {title}
  {\bibinfo {title} {Revealing the thermal properties of superconducting
  magic-angle twisted bilayer graphene},\ }\href
  {https://doi.org/10.1021/acs.nanolett.1c04512} {\bibfield  {journal}
  {\bibinfo  {journal} {Nano Letters}\ }\textbf {\bibinfo {volume} {22}},\
  \bibinfo {pages} {6465} (\bibinfo {year} {2022})}\BibitemShut {NoStop}%
\bibitem [{\citenamefont {Lian}\ \emph {et~al.}(2021)\citenamefont {Lian},
  \citenamefont {Song}, \citenamefont {Regnault}, \citenamefont {Efetov},
  \citenamefont {Yazdani},\ and\ \citenamefont
  {Bernevig}}]{lianTwisted2021tbg4}%
  \BibitemOpen
  \bibfield  {author} {\bibinfo {author} {\bibfnamefont {B.}~\bibnamefont
  {Lian}}, \bibinfo {author} {\bibfnamefont {Z.-D.}\ \bibnamefont {Song}},
  \bibinfo {author} {\bibfnamefont {N.}~\bibnamefont {Regnault}}, \bibinfo
  {author} {\bibfnamefont {D.~K.}\ \bibnamefont {Efetov}}, \bibinfo {author}
  {\bibfnamefont {A.}~\bibnamefont {Yazdani}},\ and\ \bibinfo {author}
  {\bibfnamefont {B.~A.}\ \bibnamefont {Bernevig}},\ }\bibfield  {title}
  {\bibinfo {title} {Twisted bilayer graphene. iv. exact insulator ground
  states and phase diagram},\ }\href
  {https://doi.org/10.1103/PhysRevB.103.205414} {\bibfield  {journal} {\bibinfo
   {journal} {Phys. Rev. B}\ }\textbf {\bibinfo {volume} {103}},\ \bibinfo
  {pages} {205414} (\bibinfo {year} {2021})}\BibitemShut {NoStop}%
\bibitem [{\citenamefont {Xie}\ \emph {et~al.}(2021)\citenamefont {Xie},
  \citenamefont {Cowsik}, \citenamefont {Song}, \citenamefont {Lian},
  \citenamefont {Bernevig},\ and\ \citenamefont {Regnault}}]{xieTwisted2021}%
  \BibitemOpen
  \bibfield  {author} {\bibinfo {author} {\bibfnamefont {F.}~\bibnamefont
  {Xie}}, \bibinfo {author} {\bibfnamefont {A.}~\bibnamefont {Cowsik}},
  \bibinfo {author} {\bibfnamefont {Z.-D.}\ \bibnamefont {Song}}, \bibinfo
  {author} {\bibfnamefont {B.}~\bibnamefont {Lian}}, \bibinfo {author}
  {\bibfnamefont {B.~A.}\ \bibnamefont {Bernevig}},\ and\ \bibinfo {author}
  {\bibfnamefont {N.}~\bibnamefont {Regnault}},\ }\bibfield  {title} {\bibinfo
  {title} {Twisted bilayer graphene. vi. an exact diagonalization study at
  nonzero integer filling},\ }\href
  {https://doi.org/10.1103/PhysRevB.103.205416} {\bibfield  {journal} {\bibinfo
   {journal} {Phys. Rev. B}\ }\textbf {\bibinfo {volume} {103}},\ \bibinfo
  {pages} {205416} (\bibinfo {year} {2021})}\BibitemShut {NoStop}%
\bibitem [{\citenamefont {Bultinck}\ \emph {et~al.}(2020)\citenamefont
  {Bultinck}, \citenamefont {Khalaf}, \citenamefont {Liu}, \citenamefont
  {Chatterjee}, \citenamefont {Vishwanath},\ and\ \citenamefont
  {Zaletel}}]{bultinck2020ground}%
  \BibitemOpen
  \bibfield  {author} {\bibinfo {author} {\bibfnamefont {N.}~\bibnamefont
  {Bultinck}}, \bibinfo {author} {\bibfnamefont {E.}~\bibnamefont {Khalaf}},
  \bibinfo {author} {\bibfnamefont {S.}~\bibnamefont {Liu}}, \bibinfo {author}
  {\bibfnamefont {S.}~\bibnamefont {Chatterjee}}, \bibinfo {author}
  {\bibfnamefont {A.}~\bibnamefont {Vishwanath}},\ and\ \bibinfo {author}
  {\bibfnamefont {M.~P.}\ \bibnamefont {Zaletel}},\ }\bibfield  {title}
  {\bibinfo {title} {Ground state and hidden symmetry of magic-angle graphene
  at even integer filling},\ }\href
  {https://doi.org/10.1103/PhysRevX.10.031034} {\bibfield  {journal} {\bibinfo
  {journal} {Phys. Rev. X}\ }\textbf {\bibinfo {volume} {10}},\ \bibinfo
  {pages} {031034} (\bibinfo {year} {2020})}\BibitemShut {NoStop}%
\bibitem [{\citenamefont {Wilhelm}\ \emph {et~al.}(2021)\citenamefont
  {Wilhelm}, \citenamefont {Lang},\ and\ \citenamefont
  {L\"auchli}}]{wilhelmInterplay2021}%
  \BibitemOpen
  \bibfield  {author} {\bibinfo {author} {\bibfnamefont {P.}~\bibnamefont
  {Wilhelm}}, \bibinfo {author} {\bibfnamefont {T.~C.}\ \bibnamefont {Lang}},\
  and\ \bibinfo {author} {\bibfnamefont {A.~M.}\ \bibnamefont {L\"auchli}},\
  }\bibfield  {title} {\bibinfo {title} {Interplay of fractional chern
  insulator and charge density wave phases in twisted bilayer graphene},\
  }\href {https://doi.org/10.1103/PhysRevB.103.125406} {\bibfield  {journal}
  {\bibinfo  {journal} {Phys. Rev. B}\ }\textbf {\bibinfo {volume} {103}},\
  \bibinfo {pages} {125406} (\bibinfo {year} {2021})}\BibitemShut {NoStop}%
\bibitem [{\citenamefont {Xie}\ and\ \citenamefont
  {MacDonald}(2020)}]{xieNature2020}%
  \BibitemOpen
  \bibfield  {author} {\bibinfo {author} {\bibfnamefont {M.}~\bibnamefont
  {Xie}}\ and\ \bibinfo {author} {\bibfnamefont {A.~H.}\ \bibnamefont
  {MacDonald}},\ }\bibfield  {title} {\bibinfo {title} {Nature of the
  correlated insulator states in twisted bilayer graphene},\ }\href
  {https://doi.org/10.1103/PhysRevLett.124.097601} {\bibfield  {journal}
  {\bibinfo  {journal} {Phys. Rev. Lett.}\ }\textbf {\bibinfo {volume} {124}},\
  \bibinfo {pages} {097601} (\bibinfo {year} {2020})}\BibitemShut {NoStop}%
\bibitem [{\citenamefont {Zhang}\ \emph {et~al.}(2020)\citenamefont {Zhang},
  \citenamefont {Jiang}, \citenamefont {Wang},\ and\ \citenamefont
  {Zhang}}]{YiZhang2020}%
  \BibitemOpen
  \bibfield  {author} {\bibinfo {author} {\bibfnamefont {Y.}~\bibnamefont
  {Zhang}}, \bibinfo {author} {\bibfnamefont {K.}~\bibnamefont {Jiang}},
  \bibinfo {author} {\bibfnamefont {Z.}~\bibnamefont {Wang}},\ and\ \bibinfo
  {author} {\bibfnamefont {F.}~\bibnamefont {Zhang}},\ }\bibfield  {title}
  {\bibinfo {title} {Correlated insulating phases of twisted bilayer graphene
  at commensurate filling fractions: A hartree-fock study},\ }\href
  {https://doi.org/10.1103/PhysRevB.102.035136} {\bibfield  {journal} {\bibinfo
   {journal} {Phys. Rev. B}\ }\textbf {\bibinfo {volume} {102}},\ \bibinfo
  {pages} {035136} (\bibinfo {year} {2020})}\BibitemShut {NoStop}%
\bibitem [{\citenamefont {Liu}\ and\ \citenamefont
  {Dai}(2021)}]{liuTheories2021}%
  \BibitemOpen
  \bibfield  {author} {\bibinfo {author} {\bibfnamefont {J.}~\bibnamefont
  {Liu}}\ and\ \bibinfo {author} {\bibfnamefont {X.}~\bibnamefont {Dai}},\
  }\bibfield  {title} {\bibinfo {title} {Theories for the correlated insulating
  states and quantum anomalous hall effect phenomena in twisted bilayer
  graphene},\ }\href {https://doi.org/10.1103/PhysRevB.103.035427} {\bibfield
  {journal} {\bibinfo  {journal} {Phys. Rev. B}\ }\textbf {\bibinfo {volume}
  {103}},\ \bibinfo {pages} {035427} (\bibinfo {year} {2021})}\BibitemShut
  {NoStop}%
\bibitem [{\citenamefont {Hejazi}\ \emph {et~al.}(2021)\citenamefont {Hejazi},
  \citenamefont {Chen},\ and\ \citenamefont {Balents}}]{hejaziHybrid2021}%
  \BibitemOpen
  \bibfield  {author} {\bibinfo {author} {\bibfnamefont {K.}~\bibnamefont
  {Hejazi}}, \bibinfo {author} {\bibfnamefont {X.}~\bibnamefont {Chen}},\ and\
  \bibinfo {author} {\bibfnamefont {L.}~\bibnamefont {Balents}},\ }\bibfield
  {title} {\bibinfo {title} {Hybrid wannier chern bands in magic angle twisted
  bilayer graphene and the quantized anomalous hall effect},\ }\href
  {https://doi.org/10.1103/PhysRevResearch.3.013242} {\bibfield  {journal}
  {\bibinfo  {journal} {Phys. Rev. Res.}\ }\textbf {\bibinfo {volume} {3}},\
  \bibinfo {pages} {013242} (\bibinfo {year} {2021})}\BibitemShut {NoStop}%
\bibitem [{\citenamefont {Xie}\ \emph {et~al.}(2023)\citenamefont {Xie},
  \citenamefont {Kang}, \citenamefont {Bernevig}, \citenamefont {Vafek},\ and\
  \citenamefont {Regnault}}]{xiePhase2023}%
  \BibitemOpen
  \bibfield  {author} {\bibinfo {author} {\bibfnamefont {F.}~\bibnamefont
  {Xie}}, \bibinfo {author} {\bibfnamefont {J.}~\bibnamefont {Kang}}, \bibinfo
  {author} {\bibfnamefont {B.~A.}\ \bibnamefont {Bernevig}}, \bibinfo {author}
  {\bibfnamefont {O.}~\bibnamefont {Vafek}},\ and\ \bibinfo {author}
  {\bibfnamefont {N.}~\bibnamefont {Regnault}},\ }\bibfield  {title} {\bibinfo
  {title} {Phase diagram of twisted bilayer graphene at filling factor
  $\ensuremath{\nu}=\ifmmode\pm\else\textpm\fi{}3$},\ }\href
  {https://doi.org/10.1103/PhysRevB.107.075156} {\bibfield  {journal} {\bibinfo
   {journal} {Phys. Rev. B}\ }\textbf {\bibinfo {volume} {107}},\ \bibinfo
  {pages} {075156} (\bibinfo {year} {2023})}\BibitemShut {NoStop}%
\bibitem [{\citenamefont {Lin}\ and\ \citenamefont
  {Ni}(2020)}]{linSymmetry2020}%
  \BibitemOpen
  \bibfield  {author} {\bibinfo {author} {\bibfnamefont {X.}~\bibnamefont
  {Lin}}\ and\ \bibinfo {author} {\bibfnamefont {J.}~\bibnamefont {Ni}},\
  }\bibfield  {title} {\bibinfo {title} {Symmetry breaking in the double
  moir\'e superlattices of relaxed twisted bilayer graphene on hexagonal boron
  nitride},\ }\href {https://doi.org/10.1103/PhysRevB.102.035441} {\bibfield
  {journal} {\bibinfo  {journal} {Phys. Rev. B}\ }\textbf {\bibinfo {volume}
  {102}},\ \bibinfo {pages} {035441} (\bibinfo {year} {2020})}\BibitemShut
  {NoStop}%
\bibitem [{\citenamefont {Kwan}\ \emph
  {et~al.}(2021{\natexlab{a}})\citenamefont {Kwan}, \citenamefont {Wagner},
  \citenamefont {Chakraborty}, \citenamefont {Simon},\ and\ \citenamefont
  {Parameswaran}}]{kwanDomain2021}%
  \BibitemOpen
  \bibfield  {author} {\bibinfo {author} {\bibfnamefont {Y.~H.}\ \bibnamefont
  {Kwan}}, \bibinfo {author} {\bibfnamefont {G.}~\bibnamefont {Wagner}},
  \bibinfo {author} {\bibfnamefont {N.}~\bibnamefont {Chakraborty}}, \bibinfo
  {author} {\bibfnamefont {S.~H.}\ \bibnamefont {Simon}},\ and\ \bibinfo
  {author} {\bibfnamefont {S.~A.}\ \bibnamefont {Parameswaran}},\ }\bibfield
  {title} {\bibinfo {title} {Domain wall competition in the chern insulating
  regime of twisted bilayer graphene},\ }\href
  {https://doi.org/10.1103/PhysRevB.104.115404} {\bibfield  {journal} {\bibinfo
   {journal} {Phys. Rev. B}\ }\textbf {\bibinfo {volume} {104}},\ \bibinfo
  {pages} {115404} (\bibinfo {year} {2021}{\natexlab{a}})}\BibitemShut
  {NoStop}%
\bibitem [{\citenamefont {{Datta}}\ \emph {et~al.}(2023)\citenamefont
  {{Datta}}, \citenamefont {{Calder{\'o}n}}, \citenamefont {{Camjayi}},\ and\
  \citenamefont {{Bascones}}}]{dattaHeavy2023}%
  \BibitemOpen
  \bibfield  {author} {\bibinfo {author} {\bibfnamefont {A.}~\bibnamefont
  {{Datta}}}, \bibinfo {author} {\bibfnamefont {M.~J.}\ \bibnamefont
  {{Calder{\'o}n}}}, \bibinfo {author} {\bibfnamefont {A.}~\bibnamefont
  {{Camjayi}}},\ and\ \bibinfo {author} {\bibfnamefont {E.}~\bibnamefont
  {{Bascones}}},\ }\bibfield  {title} {\bibinfo {title} {{Heavy quasiparticles
  and cascades without symmetry breaking in twisted bilayer graphene}},\ }\href
  {https://doi.org/10.48550/arXiv.2301.13024} {\bibfield  {journal} {\bibinfo
  {journal} {arXiv e-prints}\ ,\ \bibinfo {eid} {arXiv:2301.13024}} (\bibinfo
  {year} {2023})},\ \Eprint {https://arxiv.org/abs/2301.13024}
  {arXiv:2301.13024 [cond-mat.str-el]} \BibitemShut {NoStop}%
\bibitem [{\citenamefont {Kang}\ and\ \citenamefont
  {Vafek}(2020)}]{kangNon-Abelian2020}%
  \BibitemOpen
  \bibfield  {author} {\bibinfo {author} {\bibfnamefont {J.}~\bibnamefont
  {Kang}}\ and\ \bibinfo {author} {\bibfnamefont {O.}~\bibnamefont {Vafek}},\
  }\bibfield  {title} {\bibinfo {title} {Non-abelian dirac node braiding and
  near-degeneracy of correlated phases at odd integer filling in magic-angle
  twisted bilayer graphene},\ }\href
  {https://doi.org/10.1103/PhysRevB.102.035161} {\bibfield  {journal} {\bibinfo
   {journal} {Phys. Rev. B}\ }\textbf {\bibinfo {volume} {102}},\ \bibinfo
  {pages} {035161} (\bibinfo {year} {2020})}\BibitemShut {NoStop}%
\bibitem [{\citenamefont {Soejima}\ \emph {et~al.}(2020)\citenamefont
  {Soejima}, \citenamefont {Parker}, \citenamefont {Bultinck}, \citenamefont
  {Hauschild},\ and\ \citenamefont {Zaletel}}]{soejimaEfficient2020}%
  \BibitemOpen
  \bibfield  {author} {\bibinfo {author} {\bibfnamefont {T.}~\bibnamefont
  {Soejima}}, \bibinfo {author} {\bibfnamefont {D.~E.}\ \bibnamefont {Parker}},
  \bibinfo {author} {\bibfnamefont {N.}~\bibnamefont {Bultinck}}, \bibinfo
  {author} {\bibfnamefont {J.}~\bibnamefont {Hauschild}},\ and\ \bibinfo
  {author} {\bibfnamefont {M.~P.}\ \bibnamefont {Zaletel}},\ }\bibfield
  {title} {\bibinfo {title} {Efficient simulation of moir\'e materials using
  the density matrix renormalization group},\ }\href
  {https://doi.org/10.1103/PhysRevB.102.205111} {\bibfield  {journal} {\bibinfo
   {journal} {Phys. Rev. B}\ }\textbf {\bibinfo {volume} {102}},\ \bibinfo
  {pages} {205111} (\bibinfo {year} {2020})}\BibitemShut {NoStop}%
\bibitem [{\citenamefont {Zhang}\ \emph {et~al.}(2021)\citenamefont {Zhang},
  \citenamefont {Pan}, \citenamefont {Zhang}, \citenamefont {Kang},\ and\
  \citenamefont {Meng}}]{zhangMomentum2021}%
  \BibitemOpen
  \bibfield  {author} {\bibinfo {author} {\bibfnamefont {X.}~\bibnamefont
  {Zhang}}, \bibinfo {author} {\bibfnamefont {G.}~\bibnamefont {Pan}}, \bibinfo
  {author} {\bibfnamefont {Y.}~\bibnamefont {Zhang}}, \bibinfo {author}
  {\bibfnamefont {J.}~\bibnamefont {Kang}},\ and\ \bibinfo {author}
  {\bibfnamefont {Z.~Y.}\ \bibnamefont {Meng}},\ }\bibfield  {title} {\bibinfo
  {title} {Momentum space quantum monte carlo on twisted bilayer graphene},\
  }\href {https://doi.org/10.1088/0256-307X/38/7/077305} {\bibfield  {journal}
  {\bibinfo  {journal} {Chinese Physics Letters}\ }\textbf {\bibinfo {volume}
  {38}},\ \bibinfo {eid} {077305} (\bibinfo {year} {2021})}\BibitemShut
  {NoStop}%
\bibitem [{\citenamefont {Pan}\ \emph {et~al.}(2022)\citenamefont {Pan},
  \citenamefont {Zhang}, \citenamefont {Li}, \citenamefont {Sun},\ and\
  \citenamefont {Meng}}]{panDynamical2022}%
  \BibitemOpen
  \bibfield  {author} {\bibinfo {author} {\bibfnamefont {G.}~\bibnamefont
  {Pan}}, \bibinfo {author} {\bibfnamefont {X.}~\bibnamefont {Zhang}}, \bibinfo
  {author} {\bibfnamefont {H.}~\bibnamefont {Li}}, \bibinfo {author}
  {\bibfnamefont {K.}~\bibnamefont {Sun}},\ and\ \bibinfo {author}
  {\bibfnamefont {Z.~Y.}\ \bibnamefont {Meng}},\ }\bibfield  {title} {\bibinfo
  {title} {Dynamical properties of collective excitations in twisted bilayer
  graphene},\ }\href {https://doi.org/10.1103/PhysRevB.105.L121110} {\bibfield
  {journal} {\bibinfo  {journal} {Phys. Rev. B}\ }\textbf {\bibinfo {volume}
  {105}},\ \bibinfo {pages} {L121110} (\bibinfo {year} {2022})}\BibitemShut
  {NoStop}%
\bibitem [{\citenamefont {Hofmann}\ \emph {et~al.}(2022)\citenamefont
  {Hofmann}, \citenamefont {Khalaf}, \citenamefont {Vishwanath}, \citenamefont
  {Berg},\ and\ \citenamefont {Lee}}]{hofmannFermionic2022}%
  \BibitemOpen
  \bibfield  {author} {\bibinfo {author} {\bibfnamefont {J.~S.}\ \bibnamefont
  {Hofmann}}, \bibinfo {author} {\bibfnamefont {E.}~\bibnamefont {Khalaf}},
  \bibinfo {author} {\bibfnamefont {A.}~\bibnamefont {Vishwanath}}, \bibinfo
  {author} {\bibfnamefont {E.}~\bibnamefont {Berg}},\ and\ \bibinfo {author}
  {\bibfnamefont {J.~Y.}\ \bibnamefont {Lee}},\ }\bibfield  {title} {\bibinfo
  {title} {Fermionic monte carlo study of a realistic model of twisted bilayer
  graphene},\ }\href {https://doi.org/10.1103/PhysRevX.12.011061} {\bibfield
  {journal} {\bibinfo  {journal} {Phys. Rev. X}\ }\textbf {\bibinfo {volume}
  {12}},\ \bibinfo {pages} {011061} (\bibinfo {year} {2022})}\BibitemShut
  {NoStop}%
\bibitem [{\citenamefont {Pan}\ \emph {et~al.}(2023)\citenamefont {Pan},
  \citenamefont {Zhang}, \citenamefont {Lu}, \citenamefont {Li}, \citenamefont
  {Chen}, \citenamefont {Sun},\ and\ \citenamefont
  {Meng}}]{panThermodynamic2023}%
  \BibitemOpen
  \bibfield  {author} {\bibinfo {author} {\bibfnamefont {G.}~\bibnamefont
  {Pan}}, \bibinfo {author} {\bibfnamefont {X.}~\bibnamefont {Zhang}}, \bibinfo
  {author} {\bibfnamefont {H.}~\bibnamefont {Lu}}, \bibinfo {author}
  {\bibfnamefont {H.}~\bibnamefont {Li}}, \bibinfo {author} {\bibfnamefont
  {B.-B.}\ \bibnamefont {Chen}}, \bibinfo {author} {\bibfnamefont
  {K.}~\bibnamefont {Sun}},\ and\ \bibinfo {author} {\bibfnamefont {Z.~Y.}\
  \bibnamefont {Meng}},\ }\bibfield  {title} {\bibinfo {title} {Thermodynamic
  characteristic for a correlated flat-band system with a quantum anomalous
  hall ground state},\ }\href {https://doi.org/10.1103/PhysRevLett.130.016401}
  {\bibfield  {journal} {\bibinfo  {journal} {Phys. Rev. Lett.}\ }\textbf
  {\bibinfo {volume} {130}},\ \bibinfo {pages} {016401} (\bibinfo {year}
  {2023})}\BibitemShut {NoStop}%
\bibitem [{\citenamefont {Zhang}\ \emph {et~al.}(2023)\citenamefont {Zhang},
  \citenamefont {Pan}, \citenamefont {Chen}, \citenamefont {Li}, \citenamefont
  {Sun},\ and\ \citenamefont {Meng}}]{zhangPolynomial2023}%
  \BibitemOpen
  \bibfield  {author} {\bibinfo {author} {\bibfnamefont {X.}~\bibnamefont
  {Zhang}}, \bibinfo {author} {\bibfnamefont {G.}~\bibnamefont {Pan}}, \bibinfo
  {author} {\bibfnamefont {B.-B.}\ \bibnamefont {Chen}}, \bibinfo {author}
  {\bibfnamefont {H.}~\bibnamefont {Li}}, \bibinfo {author} {\bibfnamefont
  {K.}~\bibnamefont {Sun}},\ and\ \bibinfo {author} {\bibfnamefont {Z.~Y.}\
  \bibnamefont {Meng}},\ }\bibfield  {title} {\bibinfo {title} {Polynomial sign
  problem and topological mott insulator in twisted bilayer graphene},\ }\href
  {https://doi.org/10.1103/PhysRevB.107.L241105} {\bibfield  {journal}
  {\bibinfo  {journal} {Phys. Rev. B}\ }\textbf {\bibinfo {volume} {107}},\
  \bibinfo {pages} {L241105} (\bibinfo {year} {2023})}\BibitemShut {NoStop}%
\bibitem [{\citenamefont {Kuzmenko}\ \emph {et~al.}(2009)\citenamefont
  {Kuzmenko}, \citenamefont {Crassee}, \citenamefont {van~der Marel},
  \citenamefont {Blake},\ and\ \citenamefont
  {Novoselov}}]{kuzmenkoDetermination2009}%
  \BibitemOpen
  \bibfield  {author} {\bibinfo {author} {\bibfnamefont {A.~B.}\ \bibnamefont
  {Kuzmenko}}, \bibinfo {author} {\bibfnamefont {I.}~\bibnamefont {Crassee}},
  \bibinfo {author} {\bibfnamefont {D.}~\bibnamefont {van~der Marel}}, \bibinfo
  {author} {\bibfnamefont {P.}~\bibnamefont {Blake}},\ and\ \bibinfo {author}
  {\bibfnamefont {K.~S.}\ \bibnamefont {Novoselov}},\ }\bibfield  {title}
  {\bibinfo {title} {Determination of the gate-tunable band gap and
  tight-binding parameters in bilayer graphene using infrared spectroscopy},\
  }\href {https://doi.org/10.1103/PhysRevB.80.165406} {\bibfield  {journal}
  {\bibinfo  {journal} {Phys. Rev. B}\ }\textbf {\bibinfo {volume} {80}},\
  \bibinfo {pages} {165406} (\bibinfo {year} {2009})}\BibitemShut {NoStop}%
\bibitem [{\citenamefont {Andrei}\ and\ \citenamefont
  {MacDonald}(2020)}]{andrei2020graphene}%
  \BibitemOpen
  \bibfield  {author} {\bibinfo {author} {\bibfnamefont {E.~Y.}\ \bibnamefont
  {Andrei}}\ and\ \bibinfo {author} {\bibfnamefont {A.~H.}\ \bibnamefont
  {MacDonald}},\ }\bibfield  {title} {\bibinfo {title} {Graphene bilayers with
  a twist},\ }\href {https://doi.org/10.1038/s41563-020-00840-0} {\bibfield
  {journal} {\bibinfo  {journal} {Nature materials}\ }\textbf {\bibinfo
  {volume} {19}},\ \bibinfo {pages} {1265} (\bibinfo {year}
  {2020})}\BibitemShut {NoStop}%
\bibitem [{\citenamefont {Xu}\ \emph {et~al.}(2020)\citenamefont {Xu},
  \citenamefont {Liu}, \citenamefont {Rhodes}, \citenamefont {Watanabe},
  \citenamefont {Taniguchi}, \citenamefont {Hone}, \citenamefont {Elser},
  \citenamefont {Mak},\ and\ \citenamefont {Shan}}]{xu2020correlated}%
  \BibitemOpen
  \bibfield  {author} {\bibinfo {author} {\bibfnamefont {Y.}~\bibnamefont
  {Xu}}, \bibinfo {author} {\bibfnamefont {S.}~\bibnamefont {Liu}}, \bibinfo
  {author} {\bibfnamefont {D.~A.}\ \bibnamefont {Rhodes}}, \bibinfo {author}
  {\bibfnamefont {K.}~\bibnamefont {Watanabe}}, \bibinfo {author}
  {\bibfnamefont {T.}~\bibnamefont {Taniguchi}}, \bibinfo {author}
  {\bibfnamefont {J.}~\bibnamefont {Hone}}, \bibinfo {author} {\bibfnamefont
  {V.}~\bibnamefont {Elser}}, \bibinfo {author} {\bibfnamefont {K.~F.}\
  \bibnamefont {Mak}},\ and\ \bibinfo {author} {\bibfnamefont {J.}~\bibnamefont
  {Shan}},\ }\bibfield  {title} {\bibinfo {title} {Correlated insulating states
  at fractional fillings of moir{\'e} superlattices},\ }\href
  {https://doi.org/10.1038/s41586-020-2868-6} {\bibfield  {journal} {\bibinfo
  {journal} {Nature}\ }\textbf {\bibinfo {volume} {587}},\ \bibinfo {pages}
  {214} (\bibinfo {year} {2020})}\BibitemShut {NoStop}%
\bibitem [{\citenamefont {Saito}\ \emph {et~al.}(2020)\citenamefont {Saito},
  \citenamefont {Ge}, \citenamefont {Watanabe}, \citenamefont {Taniguchi},\
  and\ \citenamefont {Young}}]{saito2020independent}%
  \BibitemOpen
  \bibfield  {author} {\bibinfo {author} {\bibfnamefont {Y.}~\bibnamefont
  {Saito}}, \bibinfo {author} {\bibfnamefont {J.}~\bibnamefont {Ge}}, \bibinfo
  {author} {\bibfnamefont {K.}~\bibnamefont {Watanabe}}, \bibinfo {author}
  {\bibfnamefont {T.}~\bibnamefont {Taniguchi}},\ and\ \bibinfo {author}
  {\bibfnamefont {A.~F.}\ \bibnamefont {Young}},\ }\bibfield  {title} {\bibinfo
  {title} {Independent superconductors and correlated insulators in twisted
  bilayer graphene},\ }\href {https://doi.org/10.1038/s41567-020-0928-3}
  {\bibfield  {journal} {\bibinfo  {journal} {Nature Physics}\ }\textbf
  {\bibinfo {volume} {16}},\ \bibinfo {pages} {926} (\bibinfo {year}
  {2020})}\BibitemShut {NoStop}%
\bibitem [{\citenamefont {Lu}\ \emph {et~al.}(2019)\citenamefont {Lu},
  \citenamefont {Stepanov}, \citenamefont {Yang}, \citenamefont {Xie},
  \citenamefont {Aamir}, \citenamefont {Das}, \citenamefont {Urgell},
  \citenamefont {Watanabe}, \citenamefont {Taniguchi}, \citenamefont {Zhang}
  \emph {et~al.}}]{luSuperconductors2019}%
  \BibitemOpen
  \bibfield  {author} {\bibinfo {author} {\bibfnamefont {X.}~\bibnamefont
  {Lu}}, \bibinfo {author} {\bibfnamefont {P.}~\bibnamefont {Stepanov}},
  \bibinfo {author} {\bibfnamefont {W.}~\bibnamefont {Yang}}, \bibinfo {author}
  {\bibfnamefont {M.}~\bibnamefont {Xie}}, \bibinfo {author} {\bibfnamefont
  {M.~A.}\ \bibnamefont {Aamir}}, \bibinfo {author} {\bibfnamefont
  {I.}~\bibnamefont {Das}}, \bibinfo {author} {\bibfnamefont {C.}~\bibnamefont
  {Urgell}}, \bibinfo {author} {\bibfnamefont {K.}~\bibnamefont {Watanabe}},
  \bibinfo {author} {\bibfnamefont {T.}~\bibnamefont {Taniguchi}}, \bibinfo
  {author} {\bibfnamefont {G.}~\bibnamefont {Zhang}}, \emph {et~al.},\
  }\bibfield  {title} {\bibinfo {title} {Superconductors, orbital magnets and
  correlated states in magic-angle bilayer graphene},\ }\href
  {https://doi.org/10.1038/s41586-019-1695-0} {\bibfield  {journal} {\bibinfo
  {journal} {Nature}\ }\textbf {\bibinfo {volume} {574}},\ \bibinfo {pages}
  {653} (\bibinfo {year} {2019})}\BibitemShut {NoStop}%
\bibitem [{\citenamefont {Stepanov}\ \emph {et~al.}(2020)\citenamefont
  {Stepanov}, \citenamefont {Das}, \citenamefont {Lu}, \citenamefont
  {Fahimniya}, \citenamefont {Watanabe}, \citenamefont {Taniguchi},
  \citenamefont {Koppens}, \citenamefont {Lischner}, \citenamefont {Levitov},\
  and\ \citenamefont {Efetov}}]{stepanov2020untying}%
  \BibitemOpen
  \bibfield  {author} {\bibinfo {author} {\bibfnamefont {P.}~\bibnamefont
  {Stepanov}}, \bibinfo {author} {\bibfnamefont {I.}~\bibnamefont {Das}},
  \bibinfo {author} {\bibfnamefont {X.}~\bibnamefont {Lu}}, \bibinfo {author}
  {\bibfnamefont {A.}~\bibnamefont {Fahimniya}}, \bibinfo {author}
  {\bibfnamefont {K.}~\bibnamefont {Watanabe}}, \bibinfo {author}
  {\bibfnamefont {T.}~\bibnamefont {Taniguchi}}, \bibinfo {author}
  {\bibfnamefont {F.~H.}\ \bibnamefont {Koppens}}, \bibinfo {author}
  {\bibfnamefont {J.}~\bibnamefont {Lischner}}, \bibinfo {author}
  {\bibfnamefont {L.}~\bibnamefont {Levitov}},\ and\ \bibinfo {author}
  {\bibfnamefont {D.~K.}\ \bibnamefont {Efetov}},\ }\bibfield  {title}
  {\bibinfo {title} {Untying the insulating and superconducting orders in
  magic-angle graphene},\ }\href {https://doi.org/10.1038/s41586-020-2459-6}
  {\bibfield  {journal} {\bibinfo  {journal} {Nature}\ }\textbf {\bibinfo
  {volume} {583}},\ \bibinfo {pages} {375} (\bibinfo {year}
  {2020})}\BibitemShut {NoStop}%
\bibitem [{\citenamefont {Nam}\ and\ \citenamefont
  {Koshino}(2017)}]{namLattice2017}%
  \BibitemOpen
  \bibfield  {author} {\bibinfo {author} {\bibfnamefont {N.~N.~T.}\
  \bibnamefont {Nam}}\ and\ \bibinfo {author} {\bibfnamefont {M.}~\bibnamefont
  {Koshino}},\ }\bibfield  {title} {\bibinfo {title} {Lattice relaxation and
  energy band modulation in twisted bilayer graphene},\ }\href
  {https://doi.org/10.1103/PhysRevB.96.075311} {\bibfield  {journal} {\bibinfo
  {journal} {Phys. Rev. B}\ }\textbf {\bibinfo {volume} {96}},\ \bibinfo
  {pages} {075311} (\bibinfo {year} {2017})}\BibitemShut {NoStop}%
\bibitem [{\citenamefont {Koshino}\ \emph {et~al.}(2018)\citenamefont
  {Koshino}, \citenamefont {Yuan}, \citenamefont {Koretsune}, \citenamefont
  {Ochi}, \citenamefont {Kuroki},\ and\ \citenamefont
  {Fu}}]{koshinoMaximally2018}%
  \BibitemOpen
  \bibfield  {author} {\bibinfo {author} {\bibfnamefont {M.}~\bibnamefont
  {Koshino}}, \bibinfo {author} {\bibfnamefont {N.~F.~Q.}\ \bibnamefont
  {Yuan}}, \bibinfo {author} {\bibfnamefont {T.}~\bibnamefont {Koretsune}},
  \bibinfo {author} {\bibfnamefont {M.}~\bibnamefont {Ochi}}, \bibinfo {author}
  {\bibfnamefont {K.}~\bibnamefont {Kuroki}},\ and\ \bibinfo {author}
  {\bibfnamefont {L.}~\bibnamefont {Fu}},\ }\bibfield  {title} {\bibinfo
  {title} {Maximally localized wannier orbitals and the extended hubbard model
  for twisted bilayer graphene},\ }\href
  {https://doi.org/10.1103/PhysRevX.8.031087} {\bibfield  {journal} {\bibinfo
  {journal} {Phys. Rev. X}\ }\textbf {\bibinfo {volume} {8}},\ \bibinfo {pages}
  {031087} (\bibinfo {year} {2018})}\BibitemShut {NoStop}%
\bibitem [{\citenamefont {Xi}(2022)}]{xi2022quantum}%
  \BibitemOpen
  \bibfield  {author} {\bibinfo {author} {\bibfnamefont {D.}~\bibnamefont
  {Xi}},\ }\bibfield  {title} {\bibinfo {title} {Quantum monte carlo
  simulations in momentum space},\ }\href
  {https://doi.org/10.1088/0256-307X/39/5/050101} {\bibfield  {journal}
  {\bibinfo  {journal} {Chinese Physics Letters}\ }\textbf {\bibinfo {volume}
  {39}},\ \bibinfo {pages} {50101} (\bibinfo {year} {2022})}\BibitemShut
  {NoStop}%
\bibitem [{\citenamefont {Li}\ \emph {et~al.}(2021)\citenamefont {Li},
  \citenamefont {Kumar}, \citenamefont {Sun},\ and\ \citenamefont
  {Lin}}]{liSpontaneous2021}%
  \BibitemOpen
  \bibfield  {author} {\bibinfo {author} {\bibfnamefont {H.}~\bibnamefont
  {Li}}, \bibinfo {author} {\bibfnamefont {U.}~\bibnamefont {Kumar}}, \bibinfo
  {author} {\bibfnamefont {K.}~\bibnamefont {Sun}},\ and\ \bibinfo {author}
  {\bibfnamefont {S.-Z.}\ \bibnamefont {Lin}},\ }\bibfield  {title} {\bibinfo
  {title} {Spontaneous fractional chern insulators in transition metal
  dichalcogenide moir\'e superlattices},\ }\href
  {https://doi.org/10.1103/PhysRevResearch.3.L032070} {\bibfield  {journal}
  {\bibinfo  {journal} {Phys. Rev. Research}\ }\textbf {\bibinfo {volume}
  {3}},\ \bibinfo {pages} {L032070} (\bibinfo {year} {2021})}\BibitemShut
  {NoStop}%
\bibitem [{\citenamefont {Song}\ and\ \citenamefont
  {Bernevig}(2022)}]{songMagic-Angle2022}%
  \BibitemOpen
  \bibfield  {author} {\bibinfo {author} {\bibfnamefont {Z.-D.}\ \bibnamefont
  {Song}}\ and\ \bibinfo {author} {\bibfnamefont {B.~A.}\ \bibnamefont
  {Bernevig}},\ }\bibfield  {title} {\bibinfo {title} {Magic-angle twisted
  bilayer graphene as a topological heavy fermion problem},\ }\href
  {https://doi.org/10.1103/PhysRevLett.129.047601} {\bibfield  {journal}
  {\bibinfo  {journal} {Phys. Rev. Lett.}\ }\textbf {\bibinfo {volume} {129}},\
  \bibinfo {pages} {047601} (\bibinfo {year} {2022})}\BibitemShut {NoStop}%
\bibitem [{\citenamefont {Shi}\ and\ \citenamefont
  {Dai}(2022)}]{shiHeavey-fermion2022}%
  \BibitemOpen
  \bibfield  {author} {\bibinfo {author} {\bibfnamefont {H.}~\bibnamefont
  {Shi}}\ and\ \bibinfo {author} {\bibfnamefont {X.}~\bibnamefont {Dai}},\
  }\bibfield  {title} {\bibinfo {title} {Heavy-fermion representation for
  twisted bilayer graphene systems},\ }\href
  {https://doi.org/10.1103/PhysRevB.106.245129} {\bibfield  {journal} {\bibinfo
   {journal} {Phys. Rev. B}\ }\textbf {\bibinfo {volume} {106}},\ \bibinfo
  {pages} {245129} (\bibinfo {year} {2022})}\BibitemShut {NoStop}%
\bibitem [{\citenamefont {Călugăru}\ \emph {et~al.}(2023)\citenamefont
  {Călugăru}, \citenamefont {Borovkov}, \citenamefont {Lau}, \citenamefont
  {Coleman}, \citenamefont {Song},\ and\ \citenamefont
  {Bernevig}}]{calugaruTBG2023}%
  \BibitemOpen
  \bibfield  {author} {\bibinfo {author} {\bibfnamefont {D.}~\bibnamefont
  {Călugăru}}, \bibinfo {author} {\bibfnamefont {M.}~\bibnamefont
  {Borovkov}}, \bibinfo {author} {\bibfnamefont {L.~L.}\ \bibnamefont {Lau}},
  \bibinfo {author} {\bibfnamefont {P.}~\bibnamefont {Coleman}}, \bibinfo
  {author} {\bibfnamefont {Z.-D.}\ \bibnamefont {Song}},\ and\ \bibinfo
  {author} {\bibfnamefont {B.~A.}\ \bibnamefont {Bernevig}},\ }\bibfield
  {title} {\bibinfo {title} {Tbg as topological heavy fermion: Ii. analytical
  approximations of the model parameters},\ }\href
  {https://doi.org/10.48550/arXiv.2303.03429} {\bibfield  {journal} {\bibinfo
  {journal} {arXiv preprint arXiv:2303.03429}\ } (\bibinfo {year}
  {2023})}\BibitemShut {NoStop}%
\bibitem [{\citenamefont {Hu}\ \emph {et~al.}(2023{\natexlab{a}})\citenamefont
  {Hu}, \citenamefont {Bernevig},\ and\ \citenamefont {Tsvelik}}]{huKondo2023}%
  \BibitemOpen
  \bibfield  {author} {\bibinfo {author} {\bibfnamefont {H.}~\bibnamefont
  {Hu}}, \bibinfo {author} {\bibfnamefont {B.~A.}\ \bibnamefont {Bernevig}},\
  and\ \bibinfo {author} {\bibfnamefont {A.~M.}\ \bibnamefont {Tsvelik}},\
  }\bibfield  {title} {\bibinfo {title} {Kondo lattice model of magic-angle
  twisted-bilayer graphene: Hund's rule, local-moment fluctuations, and
  low-energy effective theory},\ }\href
  {https://doi.org/10.1103/PhysRevLett.131.026502} {\bibfield  {journal}
  {\bibinfo  {journal} {Phys. Rev. Lett.}\ }\textbf {\bibinfo {volume} {131}},\
  \bibinfo {pages} {026502} (\bibinfo {year} {2023}{\natexlab{a}})}\BibitemShut
  {NoStop}%
\bibitem [{\citenamefont {Hu}\ \emph {et~al.}(2023{\natexlab{b}})\citenamefont
  {Hu}, \citenamefont {Rai}, \citenamefont {Crippa}, \citenamefont
  {Herzog-Arbeitman}, \citenamefont {Călugăru}, \citenamefont {Wehling},
  \citenamefont {Sangiovanni}, \citenamefont {Valenti}, \citenamefont
  {Tsvelik},\ and\ \citenamefont {Bernevig}}]{huSymmetric2023}%
  \BibitemOpen
  \bibfield  {author} {\bibinfo {author} {\bibfnamefont {H.}~\bibnamefont
  {Hu}}, \bibinfo {author} {\bibfnamefont {G.}~\bibnamefont {Rai}}, \bibinfo
  {author} {\bibfnamefont {L.}~\bibnamefont {Crippa}}, \bibinfo {author}
  {\bibfnamefont {J.}~\bibnamefont {Herzog-Arbeitman}}, \bibinfo {author}
  {\bibfnamefont {D.}~\bibnamefont {Călugăru}}, \bibinfo {author}
  {\bibfnamefont {T.}~\bibnamefont {Wehling}}, \bibinfo {author} {\bibfnamefont
  {G.}~\bibnamefont {Sangiovanni}}, \bibinfo {author} {\bibfnamefont
  {R.}~\bibnamefont {Valenti}}, \bibinfo {author} {\bibfnamefont {A.~M.}\
  \bibnamefont {Tsvelik}},\ and\ \bibinfo {author} {\bibfnamefont {B.~A.}\
  \bibnamefont {Bernevig}},\ }\bibfield  {title} {\bibinfo {title} {Symmetric
  kondo lattice states in doped strained twisted bilayer graphene},\ }\href
  {https://doi.org/10.48550/arXiv.2301.04673} {\bibfield  {journal} {\bibinfo
  {journal} {arXiv preprint arXiv:2301.04673}\ } (\bibinfo {year}
  {2023}{\natexlab{b}})}\BibitemShut {NoStop}%
\bibitem [{\citenamefont {Chou}\ and\ \citenamefont
  {Das~Sarma}(2023{\natexlab{a}})}]{chouKondo2023}%
  \BibitemOpen
  \bibfield  {author} {\bibinfo {author} {\bibfnamefont {Y.-Z.}\ \bibnamefont
  {Chou}}\ and\ \bibinfo {author} {\bibfnamefont {S.}~\bibnamefont
  {Das~Sarma}},\ }\bibfield  {title} {\bibinfo {title} {Kondo lattice model in
  magic-angle twisted bilayer graphene},\ }\href
  {https://doi.org/10.1103/PhysRevLett.131.026501} {\bibfield  {journal}
  {\bibinfo  {journal} {Phys. Rev. Lett.}\ }\textbf {\bibinfo {volume} {131}},\
  \bibinfo {pages} {026501} (\bibinfo {year} {2023}{\natexlab{a}})}\BibitemShut
  {NoStop}%
\bibitem [{\citenamefont {Chou}\ and\ \citenamefont
  {Das~Sarma}(2023{\natexlab{b}})}]{chouScaling2023}%
  \BibitemOpen
  \bibfield  {author} {\bibinfo {author} {\bibfnamefont {Y.}~\bibnamefont
  {Chou}}\ and\ \bibinfo {author} {\bibfnamefont {S.}~\bibnamefont
  {Das~Sarma}},\ }\bibfield  {title} {\bibinfo {title} {Scaling theory of
  intrinsic kondo and hund's rule interactions in magic-angle twisted bilayer
  graphene},\ }\href {https://doi.org/10.48550/arXiv.2306.03121} {\bibfield
  {journal} {\bibinfo  {journal} {arXiv preprint arXiv:2306.03121}\ } (\bibinfo
  {year} {2023}{\natexlab{b}})}\BibitemShut {NoStop}%
\bibitem [{\citenamefont {Liu}\ \emph {et~al.}(2021{\natexlab{b}})\citenamefont
  {Liu}, \citenamefont {Khalaf}, \citenamefont {Lee},\ and\ \citenamefont
  {Vishwanath}}]{liuNematic2021}%
  \BibitemOpen
  \bibfield  {author} {\bibinfo {author} {\bibfnamefont {S.}~\bibnamefont
  {Liu}}, \bibinfo {author} {\bibfnamefont {E.}~\bibnamefont {Khalaf}},
  \bibinfo {author} {\bibfnamefont {J.~Y.}\ \bibnamefont {Lee}},\ and\ \bibinfo
  {author} {\bibfnamefont {A.}~\bibnamefont {Vishwanath}},\ }\bibfield  {title}
  {\bibinfo {title} {Nematic topological semimetal and insulator in magic-angle
  bilayer graphene at charge neutrality},\ }\href
  {https://doi.org/10.1103/PhysRevResearch.3.013033} {\bibfield  {journal}
  {\bibinfo  {journal} {Phys. Rev. Research}\ }\textbf {\bibinfo {volume}
  {3}},\ \bibinfo {pages} {013033} (\bibinfo {year}
  {2021}{\natexlab{b}})}\BibitemShut {NoStop}%
\bibitem [{\citenamefont {Nuckolls}\ \emph {et~al.}(2023)\citenamefont
  {Nuckolls}, \citenamefont {Lee}, \citenamefont {Oh}, \citenamefont {Wong},
  \citenamefont {Soejima}, \citenamefont {Hong}, \citenamefont
  {C{\u{a}}lug{\u{a}}ru}, \citenamefont {Herzog-Arbeitman}, \citenamefont
  {Bernevig}, \citenamefont {Watanabe} \emph {et~al.}}]{nuckolls2023quantum}%
  \BibitemOpen
  \bibfield  {author} {\bibinfo {author} {\bibfnamefont {K.~P.}\ \bibnamefont
  {Nuckolls}}, \bibinfo {author} {\bibfnamefont {R.~L.}\ \bibnamefont {Lee}},
  \bibinfo {author} {\bibfnamefont {M.}~\bibnamefont {Oh}}, \bibinfo {author}
  {\bibfnamefont {D.}~\bibnamefont {Wong}}, \bibinfo {author} {\bibfnamefont
  {T.}~\bibnamefont {Soejima}}, \bibinfo {author} {\bibfnamefont {J.~P.}\
  \bibnamefont {Hong}}, \bibinfo {author} {\bibfnamefont {D.}~\bibnamefont
  {C{\u{a}}lug{\u{a}}ru}}, \bibinfo {author} {\bibfnamefont {J.}~\bibnamefont
  {Herzog-Arbeitman}}, \bibinfo {author} {\bibfnamefont {B.~A.}\ \bibnamefont
  {Bernevig}}, \bibinfo {author} {\bibfnamefont {K.}~\bibnamefont {Watanabe}},
  \emph {et~al.},\ }\bibfield  {title} {\bibinfo {title} {Quantum textures of
  the many-body wavefunctions in magic-angle graphene},\ }\href
  {https://doi.org/10.1038/s41586-023-06226-x} {\bibfield  {journal} {\bibinfo
  {journal} {Nature}\ }\textbf {\bibinfo {volume} {620}},\ \bibinfo {pages}
  {525} (\bibinfo {year} {2023})}\BibitemShut {NoStop}%
\bibitem [{\citenamefont {Hong}\ \emph {et~al.}(2022)\citenamefont {Hong},
  \citenamefont {Soejima},\ and\ \citenamefont {Zaletel}}]{Hong2022Detecting}%
  \BibitemOpen
  \bibfield  {author} {\bibinfo {author} {\bibfnamefont {J.~P.}\ \bibnamefont
  {Hong}}, \bibinfo {author} {\bibfnamefont {T.}~\bibnamefont {Soejima}},\ and\
  \bibinfo {author} {\bibfnamefont {M.~P.}\ \bibnamefont {Zaletel}},\
  }\bibfield  {title} {\bibinfo {title} {Detecting symmetry breaking in magic
  angle graphene using scanning tunneling microscopy},\ }\href
  {https://doi.org/10.1103/PhysRevLett.129.147001} {\bibfield  {journal}
  {\bibinfo  {journal} {Phys. Rev. Lett.}\ }\textbf {\bibinfo {volume} {129}},\
  \bibinfo {pages} {147001} (\bibinfo {year} {2022})}\BibitemShut {NoStop}%
\bibitem [{\citenamefont {Kwan}\ \emph
  {et~al.}(2021{\natexlab{b}})\citenamefont {Kwan}, \citenamefont {Wagner},
  \citenamefont {Soejima}, \citenamefont {Zaletel}, \citenamefont {Simon},
  \citenamefont {Parameswaran},\ and\ \citenamefont
  {Bultinck}}]{Kwan2021Kekule}%
  \BibitemOpen
  \bibfield  {author} {\bibinfo {author} {\bibfnamefont {Y.~H.}\ \bibnamefont
  {Kwan}}, \bibinfo {author} {\bibfnamefont {G.}~\bibnamefont {Wagner}},
  \bibinfo {author} {\bibfnamefont {T.}~\bibnamefont {Soejima}}, \bibinfo
  {author} {\bibfnamefont {M.~P.}\ \bibnamefont {Zaletel}}, \bibinfo {author}
  {\bibfnamefont {S.~H.}\ \bibnamefont {Simon}}, \bibinfo {author}
  {\bibfnamefont {S.~A.}\ \bibnamefont {Parameswaran}},\ and\ \bibinfo {author}
  {\bibfnamefont {N.}~\bibnamefont {Bultinck}},\ }\bibfield  {title} {\bibinfo
  {title} {Kekul\'e spiral order at all nonzero integer fillings in twisted
  bilayer graphene},\ }\href {https://doi.org/10.1103/PhysRevX.11.041063}
  {\bibfield  {journal} {\bibinfo  {journal} {Phys. Rev. X}\ }\textbf {\bibinfo
  {volume} {11}},\ \bibinfo {pages} {041063} (\bibinfo {year}
  {2021}{\natexlab{b}})}\BibitemShut {NoStop}%
\bibitem [{\citenamefont {Wang}\ \emph {et~al.}(2023)\citenamefont {Wang},
  \citenamefont {Parker}, \citenamefont {Soejima}, \citenamefont {Hauschild},
  \citenamefont {Anand}, \citenamefont {Bultinck},\ and\ \citenamefont
  {Zaletel}}]{Wang2023Ground-state}%
  \BibitemOpen
  \bibfield  {author} {\bibinfo {author} {\bibfnamefont {T.}~\bibnamefont
  {Wang}}, \bibinfo {author} {\bibfnamefont {D.~E.}\ \bibnamefont {Parker}},
  \bibinfo {author} {\bibfnamefont {T.}~\bibnamefont {Soejima}}, \bibinfo
  {author} {\bibfnamefont {J.}~\bibnamefont {Hauschild}}, \bibinfo {author}
  {\bibfnamefont {S.}~\bibnamefont {Anand}}, \bibinfo {author} {\bibfnamefont
  {N.}~\bibnamefont {Bultinck}},\ and\ \bibinfo {author} {\bibfnamefont
  {M.~P.}\ \bibnamefont {Zaletel}},\ }\bibfield  {title} {\bibinfo {title}
  {Ground-state order in magic-angle graphene at filling
  $\ensuremath{\nu}=\ensuremath{-}3$: A full-scale density matrix
  renormalization group study},\ }\href
  {https://doi.org/10.1103/PhysRevB.108.235128} {\bibfield  {journal} {\bibinfo
   {journal} {Phys. Rev. B}\ }\textbf {\bibinfo {volume} {108}},\ \bibinfo
  {pages} {235128} (\bibinfo {year} {2023})}\BibitemShut {NoStop}%
\bibitem [{\citenamefont {Chen}\ \emph {et~al.}(2021)\citenamefont {Chen},
  \citenamefont {Liao}, \citenamefont {Chen}, \citenamefont {Vafek},
  \citenamefont {Kang}, \citenamefont {Li},\ and\ \citenamefont
  {Meng}}]{chenRealization2021}%
  \BibitemOpen
  \bibfield  {author} {\bibinfo {author} {\bibfnamefont {B.-B.}\ \bibnamefont
  {Chen}}, \bibinfo {author} {\bibfnamefont {Y.~D.}\ \bibnamefont {Liao}},
  \bibinfo {author} {\bibfnamefont {Z.}~\bibnamefont {Chen}}, \bibinfo {author}
  {\bibfnamefont {O.}~\bibnamefont {Vafek}}, \bibinfo {author} {\bibfnamefont
  {J.}~\bibnamefont {Kang}}, \bibinfo {author} {\bibfnamefont {W.}~\bibnamefont
  {Li}},\ and\ \bibinfo {author} {\bibfnamefont {Z.~Y.}\ \bibnamefont {Meng}},\
  }\bibfield  {title} {\bibinfo {title} {Realization of topological mott
  insulator in a twisted bilayer graphene lattice model},\ }\href
  {https://doi.org/10.1038/s41467-021-25438-1} {\bibfield  {journal} {\bibinfo
  {journal} {Nature Communications}\ }\textbf {\bibinfo {volume} {12}},\
  \bibinfo {pages} {5480} (\bibinfo {year} {2021})}\BibitemShut {NoStop}%
\bibitem [{\citenamefont {Lin}\ \emph {et~al.}(2022)\citenamefont {Lin},
  \citenamefont {Chen}, \citenamefont {Li}, \citenamefont {Meng},\ and\
  \citenamefont {Shi}}]{linExciton2022}%
  \BibitemOpen
  \bibfield  {author} {\bibinfo {author} {\bibfnamefont {X.}~\bibnamefont
  {Lin}}, \bibinfo {author} {\bibfnamefont {B.-B.}\ \bibnamefont {Chen}},
  \bibinfo {author} {\bibfnamefont {W.}~\bibnamefont {Li}}, \bibinfo {author}
  {\bibfnamefont {Z.~Y.}\ \bibnamefont {Meng}},\ and\ \bibinfo {author}
  {\bibfnamefont {T.}~\bibnamefont {Shi}},\ }\bibfield  {title} {\bibinfo
  {title} {Exciton proliferation and fate of the topological mott insulator in
  a twisted bilayer graphene lattice model},\ }\href
  {https://doi.org/10.1103/PhysRevLett.128.157201} {\bibfield  {journal}
  {\bibinfo  {journal} {Phys. Rev. Lett.}\ }\textbf {\bibinfo {volume} {128}},\
  \bibinfo {pages} {157201} (\bibinfo {year} {2022})}\BibitemShut {NoStop}%
\bibitem [{\citenamefont {Sandvik}(1998)}]{Sandvik1998Stochastic}%
  \BibitemOpen
  \bibfield  {author} {\bibinfo {author} {\bibfnamefont {A.~W.}\ \bibnamefont
  {Sandvik}},\ }\bibfield  {title} {\bibinfo {title} {Stochastic method for
  analytic continuation of quantum monte carlo data},\ }\href
  {https://doi.org/10.1103/PhysRevB.57.10287} {\bibfield  {journal} {\bibinfo
  {journal} {Phys. Rev. B}\ }\textbf {\bibinfo {volume} {57}},\ \bibinfo
  {pages} {10287} (\bibinfo {year} {1998})}\BibitemShut {NoStop}%
\bibitem [{\citenamefont {Shao}\ and\ \citenamefont
  {Sandvik}(2023)}]{shao2023progress}%
  \BibitemOpen
  \bibfield  {author} {\bibinfo {author} {\bibfnamefont {H.}~\bibnamefont
  {Shao}}\ and\ \bibinfo {author} {\bibfnamefont {A.~W.}\ \bibnamefont
  {Sandvik}},\ }\bibfield  {title} {\bibinfo {title} {Progress on stochastic
  analytic continuation of quantum monte carlo data},\ }\href
  {https://www.sciencedirect.com/science/article/pii/S0370157322003921?via%3Dihub}
  {\bibfield  {journal} {\bibinfo  {journal} {Physics Reports}\ }\textbf
  {\bibinfo {volume} {1003}},\ \bibinfo {pages} {1} (\bibinfo {year}
  {2023})}\BibitemShut {NoStop}%
\bibitem [{\citenamefont {Nielsen}\ and\ \citenamefont
  {Ninomiya}(1981)}]{nielsen1981no}%
  \BibitemOpen
  \bibfield  {author} {\bibinfo {author} {\bibfnamefont {H.~B.}\ \bibnamefont
  {Nielsen}}\ and\ \bibinfo {author} {\bibfnamefont {M.}~\bibnamefont
  {Ninomiya}},\ }\bibfield  {title} {\bibinfo {title} {A no-go theorem for
  regularizing chiral fermions},\ }\href
  {https://doi.org/10.1016/0370-2693(81)91026-1} {\bibfield  {journal}
  {\bibinfo  {journal} {Physics Letters B}\ }\textbf {\bibinfo {volume}
  {105}},\ \bibinfo {pages} {219} (\bibinfo {year} {1981})}\BibitemShut
  {NoStop}%
\bibitem [{\citenamefont {{Yan}}\ and\ \citenamefont
  {{Meng}}(2021)}]{yanRelating2021}%
  \BibitemOpen
  \bibfield  {author} {\bibinfo {author} {\bibfnamefont {Z.}~\bibnamefont
  {{Yan}}}\ and\ \bibinfo {author} {\bibfnamefont {Z.~Y.}\ \bibnamefont
  {{Meng}}},\ }\bibfield  {title} {\bibinfo {title} {{Relating entanglement
  spectra and energy spectra via path-integral on replica manifold}},\
  }\href@noop {} {\bibfield  {journal} {\bibinfo  {journal} {arXiv e-prints}\
  ,\ \bibinfo {eid} {arXiv:2112.05886}} (\bibinfo {year} {2021})},\ \Eprint
  {https://arxiv.org/abs/2112.05886} {arXiv:2112.05886 [cond-mat.str-el]}
  \BibitemShut {NoStop}%
\bibitem [{\citenamefont {Zhang}\ \emph {et~al.}(2022)\citenamefont {Zhang},
  \citenamefont {Sun}, \citenamefont {Li}, \citenamefont {Pan},\ and\
  \citenamefont {Meng}}]{zhangSuperconductivity2021}%
  \BibitemOpen
  \bibfield  {author} {\bibinfo {author} {\bibfnamefont {X.}~\bibnamefont
  {Zhang}}, \bibinfo {author} {\bibfnamefont {K.}~\bibnamefont {Sun}}, \bibinfo
  {author} {\bibfnamefont {H.}~\bibnamefont {Li}}, \bibinfo {author}
  {\bibfnamefont {G.}~\bibnamefont {Pan}},\ and\ \bibinfo {author}
  {\bibfnamefont {Z.~Y.}\ \bibnamefont {Meng}},\ }\bibfield  {title} {\bibinfo
  {title} {Superconductivity and bosonic fluid emerging from moir\'e flat
  bands},\ }\href {https://doi.org/10.1103/PhysRevB.106.184517} {\bibfield
  {journal} {\bibinfo  {journal} {Phys. Rev. B}\ }\textbf {\bibinfo {volume}
  {106}},\ \bibinfo {pages} {184517} (\bibinfo {year} {2022})}\BibitemShut
  {NoStop}%
\bibitem [{\citenamefont {Zhou}\ \emph {et~al.}(2021)\citenamefont {Zhou},
  \citenamefont {Yan}, \citenamefont {Wu}, \citenamefont {Sun}, \citenamefont
  {Starykh},\ and\ \citenamefont {Meng}}]{zhouAmplitude2021}%
  \BibitemOpen
  \bibfield  {author} {\bibinfo {author} {\bibfnamefont {C.}~\bibnamefont
  {Zhou}}, \bibinfo {author} {\bibfnamefont {Z.}~\bibnamefont {Yan}}, \bibinfo
  {author} {\bibfnamefont {H.-Q.}\ \bibnamefont {Wu}}, \bibinfo {author}
  {\bibfnamefont {K.}~\bibnamefont {Sun}}, \bibinfo {author} {\bibfnamefont
  {O.~A.}\ \bibnamefont {Starykh}},\ and\ \bibinfo {author} {\bibfnamefont
  {Z.~Y.}\ \bibnamefont {Meng}},\ }\bibfield  {title} {\bibinfo {title}
  {Amplitude mode in quantum magnets via dimensional crossover},\ }\href
  {https://doi.org/10.1103/PhysRevLett.126.227201} {\bibfield  {journal}
  {\bibinfo  {journal} {Phys. Rev. Lett.}\ }\textbf {\bibinfo {volume} {126}},\
  \bibinfo {pages} {227201} (\bibinfo {year} {2021})}\BibitemShut {NoStop}%
\bibitem [{\citenamefont {Yan}\ \emph {et~al.}(2021)\citenamefont {Yan},
  \citenamefont {Wang}, \citenamefont {Ma}, \citenamefont {Qi},\ and\
  \citenamefont {Meng}}]{yanTopological2021}%
  \BibitemOpen
  \bibfield  {author} {\bibinfo {author} {\bibfnamefont {Z.}~\bibnamefont
  {Yan}}, \bibinfo {author} {\bibfnamefont {Y.-C.}\ \bibnamefont {Wang}},
  \bibinfo {author} {\bibfnamefont {N.}~\bibnamefont {Ma}}, \bibinfo {author}
  {\bibfnamefont {Y.}~\bibnamefont {Qi}},\ and\ \bibinfo {author}
  {\bibfnamefont {Z.~Y.}\ \bibnamefont {Meng}},\ }\bibfield  {title} {\bibinfo
  {title} {Topological phase transition and single/multi anyon dynamics of
  $z_2$ spin liquid},\ }\href {https://doi.org/10.1038/s41535-021-00338-1}
  {\bibfield  {journal} {\bibinfo  {journal} {npj Quantum Materials}\ }\textbf
  {\bibinfo {volume} {6}},\ \bibinfo {pages} {39} (\bibinfo {year}
  {2021})}\BibitemShut {NoStop}%
\end{thebibliography}%

\end{document}